# The Outer Regions of the Galactic Bulge: II – Analysis

Rodrigo A. Ibata[1,2] and Gerard F. Gilmore[1]

1 *Institute of Astronomy, Madingley Road, Cambridge CB3 0HA*
2 *Department of Geophysics and Astronomy, University of British Columbia, Vancouver, Canada (present address)*

**ABSTRACT**

Velocity, chemical abundance and spatial distribution data for some 1500 K- and M-giants in the Galactic Bulge, which are presented in an accompanying paper (Ibata & Gilmore 1995), are analysed. We provide three levels of analysis: an almost model-independent determination of basic symmetry properties of the galactic bulge; a more detailed analysis based on conservative assumptions, which investigates the basic kinematic and chemical abundance distributions in terms of a few parameters; and a more detailed investigation of possible functional forms for the density and kinematic distributions that are consistent with the data using maximum likelihood fits.

The bulge has a well-determined linear rotation curve over the range $700 {\rm pc} \leq R \leq 3500 {\rm pc}$ with amplitude $\sim 25 \, {\rm km \, s^{-1} \, kpc^{-1}}$. Several velocity dispersion models are found to fit the data. We test the oblate isotropic bulge model of Kent (1992) by integrating the kinematics predicted by that model into our Galaxy model, and comparing with our data. We find reasonable agreement except in the most distant of our fields from the Galactic centre. We do not find a significant requirement for asymmetry or a bar from our kinematics.

Our data allow a quantitative comparison between the Milky Way bulge and the spiral galaxy bulges studied by Kormendy & Illingworth (1982) over a similar Galactocentric distance range; the Galactic bulge appears to be characteristic of the extragalactic bulge population. We determine the metallicity distribution of K giants in that subset of our fields where the photometric calibration is adequate. Our metallicity index measures the Mg$b$ feature, and is calibrated against local field standards. Thus our abundance scale assumes a similar relationship between [Mg/Fe] and [Fe/H] in the bulge to that which is seen in the solar neighbourhood. Given that assumption, the mean metallicities in these fields is [Fe/H] $\approx -0.3$ and does not vary significantly over the regions investigated. This mean is very close to the mean abundance of K giants observed recently in Baade's Window (McWilliam & Rich 1994). That is, there is no detectable abundance gradient in the Galactic bulge over the galactocentric range $500 {\rm pc} \leq {\rm R_{GC}} \leq 3500 {\rm pc}$. The overall bulge abundance distribution is in good agreement with a model that combines standard solar neighbourhood determinations of the abundance distribution functions for the halo and thick disk, together with a closed-box 'simple model' of chemical evolution with yield $\sim 0.7 Z_\odot$ to describe the bulge component.

We derive the distribution function of specific angular momentum for the bulge from our data, and compare it with determinations for the halo, the thick disk and the thin disk from Wyse & Gilmore (1992). We confirm that the bulge and the halo have angular momentum distributions which are indistinguishable, as do the thick disk and the thin disk. The bulge-halo distribution is however very different from the thick disk-thin disk distribution. This is perhaps the strongest available clue to the evolutionary relationships between different Galactic structural components.

## 1 THE SPATIAL, KINEMATIC AND ABUNDANCE STRUCTURE OF THE BULGE

The Galactic bulge is a very centrally condensed Galactic structural component. Its half-light 'radius' along the minor axis is estimated variously between $\approx 200 \, {\rm pc}$ for metal rich M-giants (Frogel et al.1990) to $\approx 800 \, {\rm pc}$ for K-band surface brightness (Kent et al.1991). The most recent determination, from the DIRBE experiment on COBE, provides a scale height of some 300pc (Weiland etal 1994). For comparison, the scale height of the Galactic old disk is $\sim 300 {\rm pc}$, while Baade's Window, the innermost region where the bulge may be well-studied with optical data, is some 550pc from the Plane. Power-law $r^{-\alpha}$ fits to integrated surface brightness and to tracer object number density both require $2.6 \lesssim \alpha \lesssim 3.7$ (Frogel et al.1990, Kent et al.1991), with $\alpha \approx 1.8$ inside $|b| \lesssim 4°$. Once the disk contribution is deducted, the integrated K-band luminosity (Kent et al.1991), the IRAS point-source surface density (Harmon & Gilmore 1988, Weinberg 1992) and the DIRBE data display a flattening of $c/a \approx 0.6$. This flattening is also seen in the inner RR Lyrae system (Wesselink 1987), though the relationship if any between these very old and rather metal-poor stars and the dominant population in the central bulge, which is rather metal-rich, remains obscure.

Several lines of evidence show that the bulge is at least partly non-axisymmetric: Binney et al.(1991) demonstrated that the existence of a bar, extending out to $\approx 2.4$ kpc from the Galactic centre, explains naturally the observed kinematics of HI, CO and CS gas in the region $|\ell| < 10°$; furthermore, Blitz & Spergel (1991) showed that the $2.4\mu$m balloon data of Matsumoto et al.(1982) within $|b| < 7°$ is brighter at positive $\ell$ than at negative $\ell$, as would be expected from a barred stellar distribution. The DIRBE data also marginally favour such an asymmetry, which was suspected in the IRAS data (Harmon & Gilmore 1988). From an analysis of IRAS point sources selected from those sources with $|b| < 3°$, Weinberg (1992) finds evidence for a large scale stellar bar in the Galactic disk which he suggests extends out to $\approx 5$ kpc; however this need not be related to the other non-axisymmetric phenomena mentioned above.

Kinematic studies of candidate bulge tracers (Miras, planetary nebulae, K giants, M giants and carbon stars) show that the bulge has a central dispersion of $\approx 130$ km s$^{-1}$ which decreases outwards (see e.g. the minor axis data compiled by Tyson 1992). In a given region there is a tendency for lower abundance tracers to have a higher velocity dispersion, as predicted by dissipational collapse models, such as the well known 'ELS' model (Eggen et al.1962). However, available information on the mean rotation ($\overline{v}_\phi$) of the bulge as a function of cylindrical coordinates $(R, z)$ in the Galaxy is confusing, because different tracers, or similar tracers with different abundance or dispersion, display markedly different $\overline{v}_\phi$; a compilation of rotation data showing this effect is displayed in Figure 1 of Menzies (1990).

The shape of the bulge velocity ellipsoid has been constrained only in Baade's Window, very near the centre. Spaenhauer et al.(1992) obtained proper motions for stars in that region; their findings are consistent with the nuclear bulge having nearly isotropic velocity dispersion.

A recent high resolution spectral survey of bulge K-giants in Baade's Window (McWilliam & Rich 1994) showed that the bulge has mean metallicity [Fe/H] = $-0.25$. That work contradicts earlier findings (*e.g.* Rich 1988, Frogel & Whitford 1987, Geisler & Friel 1992, Tyson 1991) which had deduced that the Baade's Window bulge was extremely metal-rich, with a mean abundance [Fe/H] $\approx +0.30$. The existence of an abundance gradient in the inner Galaxy is currently poorly constrained partly because there is now reason to believe that the Washington photometric system, upon which many of the relevant studies have been based (Geisler & Friel 1992, Tyson 1991, Harding & Morrison 1993), suffers from calibration limitations in the relevant abundance range. Nonetheless, there is suggestive evidence for an abundance gradient in the central bulge stellar population in at least some tracers: *e.g.* infrared photometry of M giants is consistent with an abundance change of $\approx 0.4$ dex from $-3°$ to $-12°$ (Frogel et al.1990).

### 1.1 Plan of this Paper

Our aim in this paper is to present an analysis of our kinematic, abundance and photometric data at three complementary levels. First, in a model - independent way, to identify general structural properties of the Galaxy which we may be confident are reliably determined. Second, in a model-dependent way, with a small number of conservative assumptions, to provide a larger amount of detail while retaining an appreciation of the importance of the various assumptions required. Third, with increased model-dependence, to investigate the full amount of information potentially contained in the data. This provides a wealth of information, though at the expense of risking one's comprehension in a sea of parameters.

In §2 we briefly summarise the data which is presented in detail in an accompanying paper (Ibata & Gilmore 1995). §3 outlines the stellar luminosity classification and describes the implications of the reliability of this classification for the astrophysical conclusions of this paper. §4 presents the most robust and least model-dependent deductions from the data, concluding that the bulge is kinematically symmetric, in spite of the confusing presence of the Sagittarius dwarf galaxy in our data. §5 discusses our photometric star count data, together with other surface brightness measures, to deduce the most appropriate bulge and old disk spatial density distributions. Surprisingly, our data suggest a low inner disk scale height. §6 provides a kinematic analysis utilising Kent's isotropic oblate rotator dynamical model and the density distribution of §5. The agreement of our data with the predictions of this model, at least in the inner few kpc of the Galaxy, is rather good. In §7 we extend the kinematical analysis to a more wide-ranging investigation of possible kinematic models, rotation curves, dispersion profiles, and velocity ellipsoid orientations and shapes. We employ a maximum likelihood method to compare models with the data, and analyse the data both field-by-field and *en masse*. A robust conclusion is that the bulge has a linear rotation curve with well-determined amplitude over the whole range of our data. §8 takes up this robust determination, and derives the corresponding distribution function of specific angular momentum for the bulge. We compare this to similar distribution functions for other Galactic components, the halo, the thick disk and the thin disk, to show how the bulge is closely related to the halo, but not to the disks. In §9 we combine the abundance and kinematic data, and use these to isolate the abundance distribution of the bulge. We then compare this to the 'closed-box simple model' of chemical evolution, generating a rather good fit with plausible parameters. In §10 we compare our bulge with data for external galaxies, which we are able to do over a similar range of galacto-centric distances for the first time. §11 summarises and concludes.

## 2 SUMMARY OF THE DATA

In an accompanying paper, Ibata & Gilmore (1995, henceforth Paper I) present the results of an analysis of photo-metric and spectroscopic data for a large sample of bulge giants. The stellar data are presented in Table B.1 of Paper 1 in detail. Each star has a position, a colour, a velocity, a metallicity estimate, a spectral classification and an estimate of its luminosity class (dwarf or giant). A total of $\approx 1500$ stars were observed spectroscopically. In summary:

i) Five carbon stars were discovered in the spectroscopic survey. Four of these stars lie in the $\ell = 5°$ fields, are members of the Sagittarius dwarf galaxy, and are discussed further in Ibata, Gilmore & Irwin (1995; hereafter IGI).

ii) Forty-two M giants were found in the complete sample, of which 36 also lie in the $\ell = 5°$ fields. Again, they are members of the Sagittarius dwarf, and are discussed further by IGI.

iii) Some 250 K dwarfs ($\approx 20\%$ of the sample) were classified and identified both by visual inspection of the spectra, and by use of a new classification technique based on Principal Component Analysis. This technique will be discussed further elesewhere (Ibata & Irwin, in preparation).

iv) The stars of direct interest as members of the Galactic bulge, the K giants, numbered $\approx 1200$, making up $\approx 80\%$ of the observed sample.

## 3 DWARF CONTAMINATION

Since we wish to determine parametric descriptions of the kinematics of the Galactic bulge, it is important that any stars in our sample which lie in the foreground disk or the background Sagittarius dwarf galaxy are reliably identified. For present purposes, we describe such stars as 'contaminants'. Dwarf-giant luminosity classification has been carried out for all our stars. Giant carbon stars and M-giants can be classified very reliably (cf Paper I), in large part since they are apparently bright, and so have high signal-noise ratio spectra. The fainter K dwarfs may however be less reliably classified. Thus we check by testing for their effect on the abundance and kinematic distributions. Before this, however, we emphasise, as discussed in paper I, that the number of stars in our sample classified as disk K dwarfs is in excellent agreement with the number predicted from our Galaxy model. There is therefore no reason to expect that any significant error in the luminosity classification has occured.

### 3.1 Contamination in the Velocity Distributions

The reliability of K dwarf – giant discrimination significantly affects the conclusions that will be drawn from this data set. We therefore first investigate whether the objects identified as K dwarfs have an expected velocity signature. Any K dwarfs in the sample will be foreground disk stars, some 1kpc from the Sun (cf Paper I). They will therefore have a clear kinematic distribution characteristic of the old disk. The observed velocity distributions of all stars classified as dwarfs are plotted in Figure 1, where we have superimposed the expected distribution according to our galaxy model. All fits are good to better than the $1\sigma$ level, so we are confident that the classification scheme does indeed pick out dwarfs, and does not select a significant number of giants. The relationship of disk and bulge kinematics down these lines of sight, which determines the effect of any possible residual dwarf contamination in the giant sample, is discussed later.

### 3.2 Contamination in the Metallicity Distributions

The stellar sample was divided into into giants and dwarfs using a Principal Component Analysis technique (Ibata & Irwin in preparation). This classification procedure is a function of the colour of the star and a coefficient similar to its linestrength The metallicity estimates we obtained (*cf* Paper I) are the result of an interpolation in an empirical correlation between colour and linestrength, calibrated by data for many standard stars. Thus we expect our classification and metallicity estimates to be strongly correlated. We note that the 'metallicity' derived for a dwarf is *not* calibrated in this project, so that such 'metallicity' estimates do not correspond to meaningful abundances. In particular, no zero point calibration of the dwarf abundance estimates has been attempted.

We examine the shape of the (pseudo-)'metallicity' distribution of stars classified as dwarfs, so as to see the way that the K giant metallicity distributions would be contaminated by any residual mis-classification. The distributions are shown in Figure 2; objects classified as dwarfs are strongly concentrated in the region [Fe/H] > 0. Thus, any errors in luminosity classification will preferentially provide contamination to the K giant metallicity distribution also in that range. Recall also (paper I) that the metallicity distributions are reliable only in those regions where a reliable calibration to the APM photometry is available — that is, not in the fields at $(\ell = -15°, b = -12°)$ or $(\ell = 5°, b = -20°)$, (*cf* Paper I).

## 4 MODEL-INDEPENDENT DIFFERENCES IN THE VELOCITY DISTRIBUTIONS

Using the non-parametric Kolmogorov-Smirnov (KS) test we compare the observed radial velocity distributions of different parts of the Galactic bulge. A cumulative probability distribution is constructed from the radial velocity data set in each region such that the probability of the lower velocity bound is zero and the probability of the upper bound is unity (the value of these bounds will be stated later). The KS statistic $D$ is found: this is simply the maximum value of the absolute difference between the two cumulative probability distributions. The KS test allows one to calculate the probability $P(D > \text{observed})$ that $D$ should be greater than the observed value if the distributions are drawn randomly from the same parent population. If this probability is either very small or very near unity then the distributions are significantly different.

The first question to be addressed is: is there any evidence that the observed bulge-region stars are kinematically non-axisymmetric? This question can be answered directly (in a model independent way) by comparing pairs of lines of sight which are situated at points reflected symmetrically about the bulge minor axis. The regions that we can test in this way are those at $(\ell = -5°, b = -12°)$ and $(\ell = +5°, b = -12°)$. Then the above question can be stated as: if the map $v \mapsto -v$ is applied to the velocity distributions at negative longitudes, is $P(D > \text{observed})$ such that the distributions are significantly different?

The answer to this depends on the choice of velocity bounds for the KS test. Two distributions are formed by selecting stars from the ($\ell = -5°, b = -12°$) and ($\ell = +5°, b = -12°$) fields starting from a lower velocity bound of $-250\,\mathrm{km\,s^{-1}}$ and ending at an upper velocity bound $v$. We then calculate the probability $P(D > \mathrm{Observed})$, and equate this probability with the probability that an event should be found more than $x$ standard deviations from the mean of a gaussian distribution of standard deviation $\sigma$:

$$2\frac{1}{\sqrt{2\pi}} \int_{x}^{+\infty} \exp(-t^2/2\sigma^2)dt = P(D > \mathrm{Observed}). \tag{1}$$

**Figure 1.** Comparison between the expected K dwarf velocity distributions (smooth curves) and those for the stars classified as K dwarfs. The agreement between the data and the model prediction lends confidence to the classification. The lines of sight are: (a) $\ell = -25°$, $b = -12°$, (b) $\ell = -15°$, $b = -12°$, (c) $\ell = -5°$, $b = -12°$, (d) $\ell = 5°$, $b = -12°$, (e) $\ell = 5°$, $b = -15°$ and (f) $\ell = 5°$, $b = -20°$.

Thus solving for $x$, we find the probability $P(D > \mathrm{Observed})$ in units of $\sigma$; this is shown in Figure 3. Clearly, up to $v \approx 170\,\mathrm{km\,s^{-1}}$ the radial velocity distributions in the two regions are distinguishable only below the $2\sigma$ level, which is not significant, but beyond $v \approx 170\,\mathrm{km\,s^{-1}}$ they are significantly ($\approx 4.5\sigma$) different. The (statistically significant!) kinematic feature at $\approx 170\,\mathrm{km\,s^{-1}}$ in the 5° fields is due to the (hitherto unknown) Sagittarius dwarf galaxy evident in the $\ell = +5°$ fields (Ibata, Gilmore & Irwin 1994; IGI). It is the presence of the Sagittarius dwarf in our survey fields which leads to the need for the relatively complex consideration of velocity bounds in Figure 3. At velocities below $\sim 170\mathrm{km/s}$, where the data exclusively represent the Galactic bulge, the two symmetric fields do not differ significantly in their kinematics (Figure 3). Thus we conclude that, withing the precision of the present data, the bulge is kinematically symmetric.

**Figure 2.** The pseudo-'metallicity' distribution of those stars classified as dwarfs is shown for (a) $\ell = -25°$, $b = -12°$, (b) $\ell = -15°$, $b = -12°$, (c) $\ell = -5°$, $b = -12°$, (d) $\ell = 5°$, $b = -12°$, (e) $\ell = 5°$, $b = -15°$ and (f) $\ell = 5°$, $b = -20°$. These abundances are neither calibrated nor reliable. Their significance is to show that any residual dwarf contamination in the K giant star sample will have greatest effect at the high metallicity end of the K-giant distribution, for [Fe/H] $\gtrsim 0.0$.

The next question to be asked is: is there evidence for a change in kinematics with (vertical) distance from the Galactic plane in the regions investigated? This may be tested directly with the radial velocity data obtained from the three fields parallel to the Galactic minor axis, ($\ell = $

**Figure 3.** Testing for any possible difference between the ($\ell = -5°, b = -12°$) and ($\ell = +5°, b = -12°$) fields. Two cumulative distributions are formed between $-250\,\mathrm{km\,s^{-1}}$ and an upper bound $v$ from the two data sets. The KS test statistic $D$, (the maximum value of the absolute difference between the two cumulative distributions) is found. The plot shows the probability $P(D > \mathrm{Observed})$ (in units of $\sigma$ — see the text) as a function of the upper bound $v$. Note that below $v \approx 170\,\mathrm{km\,s^{-1}}$, the two distributions cannot be considered different. The velocities shown are 'Galactocentric radial velocities' – this term is explained in Paper I.

$5°, b = -12°$), ($\ell = 5°, b = -15°$) and ($\ell = 5°, b = -20°$). It is found that, constrained by any upper or lower velocity bounds, the distributions are different at less than $1.5\sigma$. Fitting a gaussian to the data in the heliocentric radial velocity range $-100\,\mathrm{km\,s^{-1}} < v < +100\,\mathrm{km\,s^{-1}}$ (so as to avoid biasing the fit with stars from the Sagittarius dwarf galaxy), we find that the mean heliocentric velocities of the maximum-likelihood gaussian fits in the fields at $b = -12°$, $b = -15°$ and $b = -20°$ are $-6.8 \pm 4.1\,\mathrm{km\,s^{-1}}$, $-6.4 \pm 5.6\,\mathrm{km\,s^{-1}}$ and $-10.7 \pm 7.0\,\mathrm{km\,s^{-1}}$ respectively. We therefore conclude that there is no evidence from the data for vertical variations in the kinematics of bulge-region stars in the range $-1.6 \gtrsim z \gtrsim -2.7\,\mathrm{kpc}$ at $\ell = 5°$. Note that this does not necessarily imply that there is no vertical variation in bulge kinematics, since we have not considered the expected kinematics of the disk, thick disk and halo populations yet.

It has been noted that bulges of external galaxies that have so called 'boxy' isophotes generally display cylindrical rotation (Kormendy & Illingworth 1982, Shaw et al.1993); this phenomenon is observed in many low luminosity bulges (Jarvis 1986). Kent et al.(1991) finds that the $2.2\mu m$ Galactic bulge emission has slightly boxy contours, which has suggested the possibility of cylindrical rotation. The recent DIRBE data however (Weiland etal 1994) show this shape to be an artefact of patchy reddening. Since both the mean velocity and the velocity distributions in our fields at $5°$ appear to be indistinguishable, over a factor of two in distance from the plane, our data are consistent with the bulge being a cylindrical rotator.

It should be noted that the results presented in this section should be considered only as suggestive. This is because we have not taken the selection function into account in our calculations: stars in the different fields were not selected in an identical way. Thus even if the kinematics of stars in two fields were intrinsically identical, the differences in the selection functions would give rise to different proportions of disk, thick disk and halo stars in the two fields, and one obtains a spurious comparison. In section 7 we shall model the kinematic distributions and amend this deficiency by taking the selection function into account. Since this sophistication in modelling will come at the cost of introducing many free parameters into the discussion, we believe this simplistic discussion valuable to complement the later complexity.

## 5 BULGE MODEL STAR COUNTS

The contributions to the total kinematics from the thin disk, thick disk and halo are kept fixed (with the exception of the old disk scale height) as given by the parameters and relations detailed in our star count model of the Galaxy. This model will be described more fully in its present form elsewhere (Gilmore, Wyse & Hewett in preparation), but is basically that described by Gilmore (1984), by Gilmore, Wyse & Kuijken (1989), and by Gilmore, King & van der Kruit (1990). The important new features here are the parameters specific to the Galactic bulge, which we outline in this section.

The luminosity function (LF) adopted for the bulge component is taken from the disk LF by Wielen (1974) with the LF for M-giants in Baade's Window by Frogel & Whitford (1987) appended on at the bright end. An arbitrary normalisation is used, because the zero-point of the bulge density law is not known.

The colour-magnitude (CM) relation used to represent the bulge is that of M67 compiled by Chiu (1980). This disk cluster is of slightly higher metallicity ([Fe/H] = $-0.04$) than the bulk of our K-giant sample (see section 9).

### 5.1 Star Count Data and the Galaxy Model

We investigate the ability of the Galaxy model to predict stellar number counts given the selection function imposed on our data. Stars were chosen for observation (*cf* Paper I) from the colour-magnitude parameter space $14.5 \lesssim \mathrm{R} \lesssim 16.5$, $\mathrm{B_J} - \mathrm{R} \gtrsim 1.4$; the expected $\mathrm{B_J} - \mathrm{R}$ distribution of these stars in the ($\ell = -5°, b = -12°$) field is shown in Figure 4, where we have superimposed the observed distribution. In this simulation, the model works acceptably well in the red (beyond $\mathrm{B_J} - \mathrm{R} = 1.4$, where our data lie), but gives a poor fit in the blue. The steep gradient on the red side of Figure 4 clearly shows that it is necessary to calibrate the zero-point of the $\mathrm{B_J}$, R photometry to better than $\approx 0.1^m$ in order to obtain star counts accurate to better than $\approx 30\%$. Recall that in the ($\ell = -15°, b = -12°$), ($\ell = -5°, b = -12°$) and ($\ell = 5°, b = -20°$) fields no reliable CCD calibration is available. Reddening errors of $\approx 0.1^m$ in the Burstein & Heiles maps will introduce similar $\approx 30\%$ errors in the starcounts. Since we have neither precise photometric data nor accurate reddening determinations in most of our fields, the data presented here are not well suited for the study of stellar number density.

### 5.2 Bulge Density Laws

Given the assumed luminosity function and colour magnitude relation for the bulge, a density relation $\nu$ must be chosen to fit the starcounts. However, we show in IGI that the stars with heliocentric radial velocities in the range $v = 140 \pm 10\,\mathrm{km\,s^{-1}}$ in the $\ell = 5°$ fields belong to the Sagittarius dwarf galaxy, a population that is not accounted for in the galaxy model; in those fields the fit of the model to the starcounts will therefore be performed on stars with heliocentric radial velocities $v < 120\,\mathrm{km\,s^{-1}}$. With this restriction, we calculate the total number of stars expected $M_i$ and observed $N_i$ in each field $i$. The zero point of the adopted bulge density relation is obtained by minimising $\sum_i (M_i - N_i)^2$.

**Figure 4.** The $B_J - R$ distribution of stars along the line of sight $(\ell = -5°, b = -12°)$ in the magnitude interval $14.5 < R < 16.5$ is shown: the 'star' symbols represent the expected distribution, while 'dots' represent the observed distribution.

Kent et al (1991) modeled $2.4\mu m$ $SPACELAB$ data (Kent et al 1992); in the regions relevant to this investigation, the 'best fit' density relation they find is $\nu \propto K_0(s/667)$ (this gives boxy projected density contours), where $K_0$ is a modified Bessel function and $s = (R^4 + (z/0.61)^4)^{1/4}$. The 'best fit' also contains a double exponential disk with a scale length $h_R = 3001\,\mathrm{pc}$ and a scale height which is a constant $h_z = 165\,\mathrm{pc}$ from $R = 0$ to $R = 5300\,\mathrm{pc}$ but beyond which it increases linearly so as to be $h_z = 247\,\mathrm{pc}$ at $R_\odot$. The principal contributors to the $2.4\mu m$ emission are late K- and M-giants (Jones et al 1981), so we convert standard K-band flux into K- and M-giant number counts using the K-band luminosity function by Jones et al (1981). The combined model for both bulge and disk works acceptably well (*cf* section 5.1); we find that the expected number counts do not differ from the observed number counts by more than 35% in any of the fields. This confirms that the data of Kent et al (1992) is not dominated by a few very bright M-giants.

Sellwood & Sanders (1988) fit a 'maximum disk' model to $2.4\mu m$ balloon data by Matsumoto et al (1982) and to $2.2\mu m$ data by Becklin & Neugebauer (1968). The bulge density law they find is $\nu \propto r^{-1.8}[1 + (r/1\,\mathrm{kpc})^{-1}]$, which clearly asymptotes to $r^{-3.7}$. We therefore also try an $r^{-3.7}$-like model; then the best fit to K-giant number density has $(s/s_0)^{-3.7}$, where $s^2 = R^2 + (z/0.6)^2$ and $s_0 = 900\,\mathrm{pc}$.

However, we find that, as long as the old disk contribution to the total K-giant number counts is less than $\approx 30\%$ ($z_h < 220\,\mathrm{pc}$), different bulge density laws — that are able to fit the number counts — do not significantly affect the bulge kinematic parameters derived below.

### 5.3 The Old Disk Scale Height

The disk components of spiral galaxies show constant vertical scale height $h_z$ independent of $R$ (see *e.g.* the chapters by van der Kruit in Gilmore, King & van der Kruit 1990). Assuming that the presence of the bulge has no effect on the disk scale height in the central regions of the Milky Way, one would expect $z_h = $ constant through the Milky Way. Then $z_h$ can be set by the solar neighbourhood value: for instance Kuijken & Gilmore 1989 fit $h_z = 249\,\mathrm{pc}$ to local K-dwarfs. Kent et al (1991) on the other hand find a best fit to $2.4\mu m$ photometry with a bulge and a disk model in which the disk scale height is constant ($z_h = 165\,\mathrm{pc}$) out to a certain radius ($R = 5300\,\mathrm{pc}$), but beyond which it increases linearly outwards so as to be $h_z = 250\,\mathrm{pc}$ at the solar neighbourhood. The obvious interpretation is that the $2.4\mu m$ data are dominated by the young disk, which has a scale height $\lesssim 150\mathrm{pc}$, and not by the old disk. We now check to see if a similar effect appears in our data.

In the regions investigated, the ratio of expected number counts between a model with $h_z = 165\,\mathrm{pc}$ and one with $h_z = 300\,\mathrm{pc}$ is $\approx 0.13$; this makes the difference between a small disk contribution and one that has, in some fields, more than twice the observed number of stars. Since we have three fields at $\ell = 5°$, we can easily compare the vertical behaviour of the disk density at $R = 700\,\mathrm{pc}$. We deduce that, if $z_h \gtrsim 300\,\mathrm{pc}$, then either the adopted Wielen (1974) luminosity function or the colour-magnitude relation of M67 are inappropriate approximations for the inner parts of the disk, or there are significant zero-point differences in the adopted photometry (*cf* Paper I) between the fields. The problems with such a large scale height are demonstrated in Figure 5. For $165 < z_h \lesssim 250\,\mathrm{pc}$, it is possible to fit the $\ell = 5°$ fields to within the $\approx 30\%$ starcount error (*cf* section 5.1); but if $z \gtrsim 220\,\mathrm{pc}$, we are unable to fit the kinematics of the field at $(\ell = -25°, b = -12°)$ at better that about the $3\sigma$ level, unless the bulge velocity dispersion in that field has the unrealistically low value of $\approx 30\,\mathrm{km\,s^{-1}}$, when the fit is acceptable at the $2\sigma$ level. However, these problems are resolved if we adopt a scale height $z_h \lesssim 200\,\mathrm{pc}$, and we can make acceptable fits to both the starcounts and kinematics in all fields.

Thus in this respect, it appears that the disk of the Milky Way is not representative of the disk populations of spiral galaxies. The extent to which this is a structural feature of the Galaxy, indicates that our survey is still oversensitive to young disk rather than old disk stars, or indicates limitations in the calibration of our photometric data remains unclear.

## 6 COMPARISON TO KENT'S 'ISOTROPIC OBLATE ROTATOR' BULGE MODEL

The only kinematic model to date that has been constructed from a fit to the large-scale mass structure of the Galactic bulge and inner Galactic disk (as obtained from $2.4\mu m$ emission with an assumption of the mass to light ratio) is that of Kent (1992). He assumes that the bulge is an axisymmetric, oblate spheroid, of constant $M/L$ and with isotropic velocity dispersion, so that the distribution function $f$ is a function only of total energy $E$ and the $z$-component of angular momentum $L_z$. The Jeans' equations then take the

**Figure 5.** Comparison between the observed velocity distributions and the Galaxy model, where we have adopted a scale height of $z_h = 300$ pc for the disk component. The lines of sight are: (a) $\ell = -25°$, $b = -12°$, (b) $\ell = -15°$, $b = -12°$, (c) $\ell = -5°$, $b = -12°$, (d) $\ell = 5°$, $b = -12°$, (e) $\ell = 5°$, $b = -15°$ and (f) $\ell = 5°$, $b = -20°$. In each panel, the dashed-single-dotted curve represents the halo, the dotted curve represents the thick disk, the dashed-triple-dotted curve represents the bulge and the dashed curve represents the disk. We have normalised the counts so as to fit the distribution in (f). The fits in (a) and (b) can be rejected at about the $5\sigma$ level, while (c) and (d) have too many stars by $\approx 70\%$ and can be rejected at the $3\sigma$ level.

form:

$$\sigma_R^2(R,z) = \sigma_z^2 = \sigma_\phi^2 = \frac{1}{\nu} \int_z^\infty \nu \frac{\partial \Psi}{\partial z} dz, \qquad (2a)$$

$$v_\phi^2(R,z) = R\frac{\partial \Psi}{\partial R} + \frac{R}{\nu}\frac{\partial(\nu\sigma_R^2)}{\partial R}, \qquad (2b)$$

where $(R, \phi, z)$ are cylindrical coordinates, $\nu$ is stellar number density, $\Psi$ is the Galactic potential, $v_\phi$ is the mean rotational velocity of the bulge and $\sigma$ is the bulge velocity dispersion.

The forces $\partial\Psi/\partial R$ and $\partial\Psi/\partial z$ are computed, using a technique described in Kent (1989), from the mass distribution of the disk and bulge, as found by Kent et al.(1991), with the assumption that $M/L = 1$. A central $10^6$ $M_\odot$ black hole is also included in the model, though is not relevant for the present study. An algorithm to calculate $\partial\Psi/\partial R$ and $\partial\Psi/\partial z$ was kindly made available to us by Steve Kent (private communication). The mean velocity and dispersion resulting from equation 2 are in good agreement with observations for $|l| \lesssim 100\,\mathrm{pc}$ and $|b| \lesssim 1\,\mathrm{kpc}$, ie, interior to almost all our data, but has hitherto not been tested in more distant regions from the Galactic centre.

The velocity distribution of the bulge population at each point along the line of sight is contructed from the estimates of $\sigma(R, z)$ and $\overline{v}_\phi$ obtained from equations 2. We use the Galaxy model to integrate disk, thick disk and halo contributions, and add Kent's model for the bulge component; this gives the distributions shown in Figure 6. According to KS tests, all the fits except that in Figure 6e are unacceptable at about the $2 - 3\sigma$ level.

Since the observed distributions are more peaked than the above model predicts, we ask the question: are these bad fits due to contamination from the local dwarf population? Objects classified as dwarfs contaminate the sample mostly at the high metallicity end, as we showed in Figure 2. To investigate this question, we apply the above kinematic model to only those stars with metallicities $[\mathrm{Fe/H}] < -0.5$. The resulting fits are shown in Figure 7; now all are acceptable at better than the $1.5\sigma$ level, except for the field at $(\ell = -25^\circ, b = -12^\circ)$, which can be discarded at the $\approx 4.5\sigma$ level. Thus Kent's model does not work well at Galactocentric distances $R \approx 3.5\,\mathrm{kpc}$. It also appears that either the model does not predict sufficient numbers of dwarf stars, the apparent excellent agreement between model prediction and classification of dwarfs noted in §3 is chance and that there is signficant residual dwarf contamination in the giant sample, or that there is a significant dependence of velocity dispersion on metallicity in the bulge population — these possibilities will be investigated more thoroughly in section 9 below.

# 7 FUNCTIONAL FITS TO BULGE KINEMATICS

## 7.1 Maximum Likelihood

When examining the models that can be allowed or refuted by the data we fix most of the relations in the galaxy model. We wish to find the most likely parameters of the remaining relations. The likelihood $L(H|D)$ of a hypothesis (or model) $H$ given data $D$ is defined to be proportional to the probability $P(D|H)$ of the data given the hypothesis (see e.g. Edwards 1992). Maximising the likelihood $L$

$$L \propto \prod_{i=1}^{n} P(D_i|H), \qquad (3)$$

is equivalent to minimising the quantity

$$s = -2 \sum_{i=1,n} \ln P(D_i|H), \qquad (4)$$

where $D = \{D_1, D_2, \ldots, D_n\}$. (It is customary to include the factor 2 in the definition of $s$ so as to emphasize the analogy with thermodynamic entropy). Given the observed $\{i = 1, \ldots, m\}$ fields containing $n_i$ stars, the statistic $s$ is found from

$$s = -2 \sum_{i=1}^{m} \sum_{j=1}^{n_i} \ln\left[\frac{G(v_j)}{T_i}\right], \qquad (5)$$

where $G$ is the expected line of sight velocity distribution, $T_i = \int G(v)dv$ and $v_j$ are the velocity values.

The statistic $s$ is minimised by varying the chosen parameters in the model with an 'amoeba' minimising algorithm from Press et al (1986).

## 7.2 Bootstrapping

With conventional Monte-Carlo simulation, one estimates the confidence limits on the parameters $\mathbf{a}_0$ fitted to a data set by finding new parameters $\mathbf{a}_i$ which are fitted to a large number $i$ of synthetic data sets. The expectation is that the probability distribution of $\mathbf{a}_i - \mathbf{a}_0$ mirrors $\mathbf{a}_i - \mathbf{a}_{\mathrm{true}}$, where $\mathbf{a}_{\mathrm{true}}$ are the parameters one would find from a hypothetical infinitely precise experiment. The synthetic data sets are created according to one's understanding of the process being investigated and the related measurement errors.

However, since it is difficult to quantify the errors in the input parameters to our starcount model, we use instead the so-called bootstrap method, which is a Monte-Carlo simulation where the synthetic data sets are created by drawing $N$ points, with replacement, from the original data set of $N$ points. Taking $i \approx 50$ synthetic data sets in each of the observed lines of sight, we sample the distribution $\mathbf{a}_i - \mathbf{a}_{\mathrm{true}}$ from the $i$ maximum likelihood fits $\mathbf{a}_i$. For a discussion of this technique, see Press et al (1986).

## 7.3 The Fits

We now investigate some possible functional forms of $\sigma(R, z)$ and $\overline{v}_\phi$ in the bulge, finding parameters that are consistent with our data. The most-likely bulge rotation relations given the data are dependent on the bulge velocity dispersion relations and upon the scale height $z_h$ of the disk component in the inner regions of the Galaxy. We argued above (in section 5.3) our preference for the values $z_h \lesssim 200\,\mathrm{pc}$. Given that this is a major source of uncertainty in the fits below, we will present three possibilities with each fit: (N) – no disk, (K) – for $z_h = 165\,\mathrm{pc}$ (as given by Kent et al.1991) – for $z_h = 200\,\mathrm{pc}$ (the largest scale height that appears to be consistent with our data). The KS test probabilities for all the fits discussed below are given in table 1; except when otherwise stated, they are all acceptable at better than the $\approx 2\sigma$ level.

Each of the observed velocity distributions has information on the kinematics of the Galaxy over a large range of Galactocentric radius. Thus the kinematic model of the Galaxy should be fit simultaneously to the velocity distributions in all of our fields. However, the maximum likelihood fits will naturally be biased towards finding parameters that fit the better sampled fields best. We therefore first investigate, in the next two paragraphs below, what can be said about linear and cylindrical rotation in the bulge, if the kinematic distributions from each field are fit independently from one another. The results of these fits can later be compared to those in which the velocity distributions are fit simultaneously.

Figure 6. Comparison between the observed velocity distributions and the expected distribution according to a model which has standard parameters for the disk, thick disk and halo populations, but whose bulge population is an oblate isotropic rotator as fit by Kent (1992). Kent's density model for the bulge and thin disk works reasonably well, giving number counts that agree with observations to $\approx 35\%$. Note that the total counts in the modeled distributions below have been normalised to the counts in the data, and not predicted. The lines of sight are: (a) $\ell = -25°$, $b = -12°$, (b) $\ell = -15°$, $b = -12°$, (c) $\ell = -5°$, $b = -12°$, (d) $\ell = 5°$, $b = -12°$, (e) $\ell = 5°$, $b = -15°$ and (f) $\ell = 5°$, $b = -20°$. The fit in (e) is acceptable at the $1\sigma$ level, though all the rest are poor, at about the $3\sigma$ level.

### 1. Oblate bulge

We investigate models in which the bulge has an oblate stellar density distribution of the form $\nu(R, z) = \nu_0(s/s_0)^{-3.7}$, where $s^2 = x^2 + y^2 + (z/0.6)^2$ and $s_0 = 900\,\mathrm{pc}$ (cf section 5.3). The KS test comparisons between kinematic distributions from data on opposite sides of the bulge strongly support this model (cf section 4).

#### 1.a. Is the bulge a linear rotator?

**Figure 7.** The velocity distributions of K giant stars with metallicities [Fe/H] < −0.5 are compared to a model similar to that shown in Figure 6, but without a local dwarf population. The lines of sight are: (a) $\ell = -25°$, $b = -12°$, (b) $\ell = -15°$, $b = -12°$, (c) $\ell = -5°$, $b = -12°$, (d) $\ell = 5°$, $b = -12°$, (e) $\ell = 5°$, $b = -15°$ and (f) $\ell = 5°$, $b = -20°$. All fits, except that in (a), are acceptable at better than the 1.5$\sigma$ level. The fit in (a) can be discarded at the $\approx 4.5\sigma$ level. This model therefore appears to work well in the central parts of the Galaxy.

The velocity distribution in each of our fields contains information about the mean azimuthal velocity and the velocity dispersion tensor over a large range of heliocentric distance. However, the largest contribution to the distribution will come from regions near the tangent point to the line of sight in the particular field under study, where the stellar number density is highest. In a small distance range near the tangent point, we can expect the mean azimuthal velocity $v_\phi$ to be a linear function of Galactocentric radius $R$, so to first order we can attempt to fit $v_\phi$ with a function $v_\phi(R) = \Omega_b R$. We proceed to fit the coefficient $\Omega_b$ independently for each field. Then, if the fitted parameters $\Omega_b$ are the same over all fields (to within the measuring errors), the bulge can indeed be said to be a linear rotator. For these fits we select only those stars classified as giants with metallicities [Fe/H] < −0.5; this should allevi-ate contamination problems with local K dwarfs (see Figure 2) and distant disk giants (whose mean metallicity we can expect to be approximately solar). We also select only the fields at $(\ell = -25°, b = -12°)$, $(\ell = -15°, b = -12°)$ and $(\ell = -5°, b = -12°)$. The fields at $(\ell = 5°, b = -15°)$ and $(\ell = 5°, b = -20°)$ are not analysed because we wish to suppress vertical terms in the fits, while the field at $(\ell = 5°, b = -12°)$ has a velocity distribution that is substantially contaminated with stars belonging to the Sagittarius dwarf galaxy (this is discussed in IGI). We will assume that the $R$ component of the velocity dispersion is almost constant in the neighbourhood of the tangent point, i.e. $\sigma_{RR}(R) \approx$ constant.

**1.a.i.** Assuming that the bulge has isotropic velocity dispersion, i.e. $\sigma_{RR} = \sigma_{\phi\phi} = \sigma_{zz}$, we find
$(\ell = -25°, b = -12°)$: $\Omega_b = 26.6 \pm 3.1 \,\mathrm{km\,s^{-1}\,kpc^{-1}}$ and $\sigma = 48 \pm 3 \,\mathrm{km\,s^{-1}}$,
$(\ell = -15°, b = -12°)$: $\Omega_b = 25.3 \pm 4.5 \,\mathrm{km\,s^{-1}\,kpc^{-1}}$ and $\sigma = 45 \pm 6 \,\mathrm{km\,s^{-1}}$,
$(\ell = -5°, b = -12°)$: $\Omega_b = 24.8 \pm 2.7 \,\mathrm{km\,s^{-1}\,kpc^{-1}}$ and $\sigma = 71 \pm 4 \,\mathrm{km\,s^{-1}}$,

**1.a.ii.** Constraining $\sigma_{RR} : \sigma_{\phi\phi} : \sigma_{zz}$ in the ratio of solar neighbourhood disk dispersions, $39 : 23 : 20$, gives:
$(\ell = -25°, b = -12°)$: $\Omega_b = 26.6 \pm 3.7 \,\mathrm{km\,s^{-1}\,kpc^{-1}}$ and $\sigma = 74 \pm 5 \,\mathrm{km\,s^{-1}}$,
$(\ell = -15°, b = -12°)$: $\Omega_b = 25.3 \pm 4.3 \,\mathrm{km\,s^{-1}\,kpc^{-1}}$ and $\sigma = 67 \pm 10 \,\mathrm{km\,s^{-1}}$,
$(\ell = -5°, b = -12°)$: $\Omega_b = 25.4 \pm 2.9 \,\mathrm{km\,s^{-1}\,kpc^{-1}}$ and $\sigma = 86 \pm 7 \,\mathrm{km\,s^{-1}}$,

**1.a.iii.** Constraining $\sigma_{RR} : \sigma_{\phi\phi} : \sigma_{zz}$ in the ratio of solar neighbourhood halo dispersions, $131 : 102 : 89$, gives:
$(\ell = -25°, b = -12°)$: $\Omega_b = 26.6 \pm 2.9 \,\mathrm{km\,s^{-1}\,kpc^{-1}}$ and $\sigma = 59 \pm 4 \,\mathrm{km\,s^{-1}}$,
$(\ell = -15°, b = -12°)$: $\Omega_b = 25.1 \pm 5.2 \,\mathrm{km\,s^{-1}\,kpc^{-1}}$ and $\sigma = 61 \pm 6 \,\mathrm{km\,s^{-1}}$,
$(\ell = -5°, b = -12°)$: $\Omega_b = 25.1 \pm 2.0 \,\mathrm{km\,s^{-1}\,kpc^{-1}}$ and $\sigma = 77 \pm 5 \,\mathrm{km\,s^{-1}}$,

Clearly, $\Omega_b$ is not very sensitive to the assumed velocity dispersion model (the detailed relation between $\Omega_b$ and $\sigma$ will be explored below). Thus to good approximation, the bulge, as sampled with our data set, is a linear rotator, whose rotation rate is $\Omega_b \approx 25 \,\mathrm{km\,s^{-1}\,kpc^{-1}}$. It is interesting to note that, extrapolated to the solar neighbourhood, this population would have a mean azimuthal velocity identical to that of the solar neighbourhood thin disk, though this is perhaps coincidental.

### 1.b. Is the bulge a cylindrical rotator?

Some calculations presented in section 4 suggested that the mean rotation of the bulge does not vary significantly with height above the Galactic plane. We now re-investigate this point, using data from the three fields at $(\ell = 5°, b = -12°)$, $(\ell = 5°, b = -15°)$ and $(\ell = 5°, b = -20°)$. We would ideally like to choose, as above, only those stars with (say) [Fe/H] $< -0.5$, that way the contribution of the local and distant disk populations to any vertical variations in the mean velocity would be reduced. Unfortunately, we have no reliable photometry and hence no reliable classifications or metallicities for stars in the field at $(\ell = 5°, b = -20°)$; however, comparing the $(\ell = 5°, b = -12°)$ field to that at $(\ell = 5°, b = -20°)$ gives a large vertical baseline, which is crucial in answering the cylindrical rotation question. Furthermore, if we accept the preliminary photometric calibrations for the $(\ell = 5°, b = -20°)$ field (cf Paper I), and choose only those stars with [Fe/H] $< -0.5$, we would obtain the velocity distributions shown in Figure 7 which have too few stars for reliable modeling.

We will fit the velocity distribution (of all the stars) of one field at a time, finding $\Omega_b$ and $\sigma$ for each field. (In the next section we will attempt to fit $\Omega_b$ and $\sigma$ simultaneously in all our fields). Again we assume that $\sigma_{RR}(R)$ is almost constant in the neighbourhood of the tangent point.

**1.b.i.** Assuming that the velocity ellipsoid is isotropic, i.e. $\sigma_{RR} = \sigma_{\phi\phi} = \sigma_{zz}$, we find
(N) $(\ell = 5°, b = -12°)$: $\Omega_b = 15 \pm 8 \,\mathrm{km\,s^{-1}\,kpc^{-1}}$ and $\sigma = 54 \pm 5 \,\mathrm{km\,s^{-1}}$,
$(\ell = 5°, b = -15°)$: $\Omega_b = 22 \pm 12 \,\mathrm{km\,s^{-1}\,kpc^{-1}}$ and $\sigma = 70 \pm 7 \,\mathrm{km\,s^{-1}}$,
$(\ell = 5°, b = -20°)$: $\Omega_b = 18 \pm 10 \,\mathrm{km\,s^{-1}\,kpc^{-1}}$ and $\sigma = 42 \pm 5 \,\mathrm{km\,s^{-1}}$.
(K) $(\ell = 5°, b = -12°)$: $\Omega_b = 14 \pm 6 \,\mathrm{km\,s^{-1}\,kpc^{-1}}$ and $\sigma = 54 \pm 5 \,\mathrm{km\,s^{-1}}$,
$(\ell = 5°, b = -15°)$: $\Omega_b = 17 \pm 11 \,\mathrm{km\,s^{-1}\,kpc^{-1}}$ and $\sigma = 67 \pm 8 \,\mathrm{km\,s^{-1}}$,
$(\ell = 5°, b = -20°)$: $\Omega_b = -7 \pm 15 \,\mathrm{km\,s^{-1}\,kpc^{-1}}$ and $\sigma = 44 \pm 6 \,\mathrm{km\,s^{-1}}$.
(L) $(\ell = 5°, b = -12°)$: $\Omega_b = 21 \pm 5 \,\mathrm{km\,s^{-1}\,kpc^{-1}}$ and $\sigma = 49 \pm 4 \,\mathrm{km\,s^{-1}}$,
$(\ell = 5°, b = -15°)$: $\Omega_b = 16 \pm 8 \,\mathrm{km\,s^{-1}\,kpc^{-1}}$ and $\sigma = 66 \pm 10 \,\mathrm{km\,s^{-1}}$,
$(\ell = 5°, b = -20°)$: $\Omega_b = -8 \pm 10 \,\mathrm{km\,s^{-1}\,kpc^{-1}}$ and $\sigma = 44 \pm 9 \,\mathrm{km\,s^{-1}}$.

Thus, with the above assumptions, the bulge may clearly be a cylindrical rotator in (N). In (K) the mean rotational velocity in the $(\ell = 5°, b = -20°)$ field is lower than that in the fields at $(\ell = 5°, b = -12°)$ and $(\ell = 5°, b = -15°)$ by $\approx 1.5$ standard deviations, so cylindrical rotation cannot be ruled out. However, cylindrical rotation in (L) can be ruled out at the $2\sigma$ level.

**1.b.ii.** With $\sigma_{RR} : \sigma_{\phi\phi} : \sigma_{zz}$ in the ratio of solar neighbourhood disk dispersions, $39 : 23 : 20$, we find:
(N) $(\ell = 5°, b = -12°)$: $\Omega_b = 20 \pm 7 \,\mathrm{km\,s^{-1}\,kpc^{-1}}$ and $\sigma = 72 \pm 5 \,\mathrm{km\,s^{-1}}$,
$(\ell = 5°, b = -15°)$: $\Omega_b = 16 \pm 6 \,\mathrm{km\,s^{-1}\,kpc^{-1}}$ and $\sigma = 84 \pm 10 \,\mathrm{km\,s^{-1}}$,
$(\ell = 5°, b = -20°)$: $\Omega_b = -7 \pm 12 \,\mathrm{km\,s^{-1}\,kpc^{-1}}$ and $\sigma = 59 \pm 7 \,\mathrm{km\,s^{-1}}$.
(K) $(\ell = 5°, b = -12°)$: $\Omega_b = 30 \pm 11 \,\mathrm{km\,s^{-1}\,kpc^{-1}}$ and $\sigma = 64 \pm 5 \,\mathrm{km\,s^{-1}}$,
$(\ell = 5°, b = -15°)$: $\Omega_b = 15 \pm 8 \,\mathrm{km\,s^{-1}\,kpc^{-1}}$ and $\sigma = 84 \pm 7 \,\mathrm{km\,s^{-1}}$,
$(\ell = 5°, b = -20°)$: $\Omega_b = 16 \pm 10 \,\mathrm{km\,s^{-1}\,kpc^{-1}}$ and $\sigma = 82 \pm 8 \,\mathrm{km\,s^{-1}}$.
(L) $(\ell = 5°, b = -12°)$: $\Omega_b = 20 \pm 7 \,\mathrm{km\,s^{-1}\,kpc^{-1}}$ and $\sigma = 58 \pm 6 \,\mathrm{km\,s^{-1}}$,
$(\ell = 5°, b = -15°)$: $\Omega_b = 14 \pm 9 \,\mathrm{km\,s^{-1}\,kpc^{-1}}$ and $\sigma = 83 \pm 8 \,\mathrm{km\,s^{-1}}$,
$(\ell = 5°, b = -20°)$: $\Omega_b = 16 \pm 9 \,\mathrm{km\,s^{-1}\,kpc^{-1}}$ and

$$\sigma = 81 \pm 15 \, \mathrm{km\, s^{-1}}.$$

With the above assumptions, cylindrical rotation is consistent (at better than the $2\sigma$ level) with scenarios (N), (K) and (L).

**1.b.iii.** While if $\sigma_{RR} : \sigma_{\phi\phi} : \sigma_{zz}$ is in the ratio of solar neighbourhood halo dispersions, $131 : 102 : 89$, then
(N) $(\ell = 5^\circ, b = -12^\circ)$: $\Omega_b = 11 \pm 5 \, \mathrm{km\, s^{-1} kpc^{-1}}$ and
$\sigma = 59 \pm 3 \, \mathrm{km\, s^{-1}}$,
$(\ell = 5^\circ, b = -15^\circ)$: $\Omega_b = 16 \pm 7 \, \mathrm{km\, s^{-1} kpc^{-1}}$ and
$\sigma = 77 \pm 4 \, \mathrm{km\, s^{-1}}$,
$(\ell = 5^\circ, b = -20^\circ)$: $\Omega_b = -7 \pm 15 \, \mathrm{km\, s^{-1} kpc^{-1}}$ and
$\sigma = 51 \pm 6 \, \mathrm{km\, s^{-1}}$.
(K) $(\ell = 5^\circ, b = -12^\circ)$: $\Omega_b = 14 \pm 6 \, \mathrm{km\, s^{-1} kpc^{-1}}$ and
$\sigma = 61 \pm 4 \, \mathrm{km\, s^{-1}}$,
$(\ell = 5^\circ, b = -15^\circ)$: $\Omega_b = 16 \pm 9 \, \mathrm{km\, s^{-1} kpc^{-1}}$ and
$\sigma = 77 \pm 5 \, \mathrm{km\, s^{-1}}$,
$(\ell = 5^\circ, b = -20^\circ)$: $\Omega_b = -7 \pm 14 \, \mathrm{km\, s^{-1} kpc^{-1}}$ and
$\sigma = 51 \pm 7 \, \mathrm{km\, s^{-1}}$.
(L) $(\ell = 5^\circ, b = -12^\circ)$: $\Omega_b = 27 \pm 4 \, \mathrm{km\, s^{-1} kpc^{-1}}$ and
$\sigma = 58 \pm 4 \, \mathrm{km\, s^{-1}}$,
$(\ell = 5^\circ, b = -15^\circ)$: $\Omega_b = 15 \pm 10 \, \mathrm{km\, s^{-1} kpc^{-1}}$ and
$\sigma = 76 \pm 6 \, \mathrm{km\, s^{-1}}$,
$(\ell = 5^\circ, b = -20^\circ)$: $\Omega_b = -7 \pm 16 \, \mathrm{km\, s^{-1} kpc^{-1}}$ and
$\sigma = 51 \pm 6 \, \mathrm{km\, s^{-1}}$.

With the above assumptions, cylindrical rotation is consistent (at better than the $2\sigma$ level) with scenarios (N), (K) and (L).

According to the above fits to our data, cylindrical rotation is consistent with our data except if the bulge velocity ellipsoid is isotropic and the disk scale height is greater than 165 pc. Cylindrical rotation in the bulge will be discussed in more detail below (where we will model the Galaxy less simplistically, dropping the assumption that the observed velocity distribution is only due to stars near the tangent point).

In the paragraphs below, we will fit models to the velocity distributions of several fields at a time (see the introduction to this section).

### 1.c. $\sigma$ isothermal, linear rotation curve

We demand that the bulge velocity dispersion $\sigma$ is not a function of position in the Galaxy, *i.e.* $\partial\sigma/\partial R = \partial\sigma/\partial z = 0$ and that it has linear rotation $v(R) = \Omega_b R$ (where $\Omega_b$ is constant). We select data from only those fields at $b = -12^\circ$ so as to suppress vertical dependence in the model fits.

**1.c.i.** Constraining $\sigma$ such that the velocity ellipsoid is everywhere isotropic, *i.e.* $\sigma_{RR} = \sigma_{\phi\phi} = \sigma_{zz}$, we find:
(N) $\Omega_b = 27 \pm 3 \, \mathrm{km\, s^{-1} kpc^{-1}}$ and $\sigma = 58 \pm 3 \, \mathrm{km\, s^{-1}}$,
(K) $\Omega_b = 27 \pm 4 \, \mathrm{km\, s^{-1} kpc^{-1}}$ and $\sigma = 57 \pm 4 \, \mathrm{km\, s^{-1}}$,
This fit can be rejected (just) at the $2\sigma$ level.
(L) $\Omega_b = 25 \pm 3 \, \mathrm{km\, s^{-1} kpc^{-1}}$ and $\sigma = 56 \pm 3 \, \mathrm{km\, s^{-1}}$.
This fit can be rejected at the $2\sigma$ level.

**1.c.ii.** Constraining $\sigma_{RR} : \sigma_{\phi\phi} : \sigma_{zz}$ in the ratio of solar neighbourhood disk dispersions, $39 : 23 : 20$, gives:
(N) $\Omega_b = 27 \pm 3 \, \mathrm{km\, s^{-1} kpc^{-1}}$ and $\sigma_{RR} = 75 \pm 5 \, \mathrm{km\, s^{-1}}$,
(K) $\Omega_b = 27 \pm 2 \, \mathrm{km\, s^{-1} kpc^{-1}}$ and $\sigma_{RR} = 74 \pm 3 \, \mathrm{km\, s^{-1}}$,
(L) $\Omega_b = 25 \pm 3 \, \mathrm{km\, s^{-1} kpc^{-1}}$ and $\sigma_{RR} = 73 \pm 4 \, \mathrm{km\, s^{-1}}$.
All three fits can be rejected at the $2\sigma$ level.

**1.c.iii.** Next we try constraining $\sigma_{RR} : \sigma_{\phi\phi} : \sigma_{zz}$ in the ratio of solar neighbourhood halo dispersions $131 : 102 : 89$. This gives:
(N) $\Omega_b = 27 \pm 3 \, \mathrm{km\, s^{-1} kpc^{-1}}$ and $\sigma_{RR} = 67 \pm 4 \, \mathrm{km\, s^{-1}}$,
This fit can be rejected at the $2\sigma$ level.
(K) $\Omega_b = 27 \pm 3 \, \mathrm{km\, s^{-1} kpc^{-1}}$ and $\sigma_{RR} = 66 \pm 4 \, \mathrm{km\, s^{-1}}$,
This fit can be rejected (just) at the $2\sigma$ level.
(L) $\Omega_b = 25 \pm 2 \, \mathrm{km\, s^{-1} kpc^{-1}}$ and $\sigma_{RR} = 65 \pm 3 \, \mathrm{km\, s^{-1}}$.

Thus our data are consistent with the bulge being a linear rotator in all the models considered above. The bulge rotation rate is well determined in all models. The velocity dispersion in these fields is much lower than that observed in the central regions of the bulge near Baade's window, where the velocity dispersion of most tracer objects is in excess of $100 \, \mathrm{km\, s^{-1}}$ (see *e.g.* the data compiled by Tyson 1992). Tyson's compilation of minor axis observations shows a clear decrease in velocity dispersion with increasing Galactocentric distance, with $\sigma_{RR}$ tailing off to $60 - 70 \, \mathrm{km\, s^{-1}}$ at $|b| \gtrsim 10$. This is consistent with the above fits. The velocity dispersion model which has the velocity ellipsoid in the shape of the solar neighbourhood disk dispersions is not consistent with the data (at the $\approx 2\sigma$ level) — though it should be noted that the other fits are only marginally better (see appendix A).

### 1.d. $\sigma$ isothermal, cylindrical rotation

Next, we demand that the bulge velocity dispersion $\sigma$ is not a function of position in the Galaxy, *i.e.* $\partial\sigma/\partial R = \partial\sigma/\partial z = 0$ and also that the bulge behaves as a cylindrical rotator, whose mean rotation increases linearly with $R$ over the region for which we have data, *i.e.* $\overline{v_\phi}(R, z) = \Omega_b R$.

**1.d.i.** Constraining $\sigma$ such that the velocity ellipsoid is everywhere isotropic, *i.e.* $\sigma_{RR} = \sigma_{\phi\phi} = \sigma_{zz}$, we find:
(N) $\Omega_b = 27 \pm 2 \, \mathrm{km\, s^{-1} kpc^{-1}}$ and $\sigma = 58 \pm 3 \, \mathrm{km\, s^{-1}}$,
(K) $\Omega_b = 26 \pm 2 \, \mathrm{km\, s^{-1} kpc^{-1}}$ and $\sigma = 58 \pm 4 \, \mathrm{km\, s^{-1}}$,
This fit can be rejected (just) at the $2\sigma$ level.
(L) $\Omega_b = 24 \pm 2 \, \mathrm{km\, s^{-1} kpc^{-1}}$ and $\sigma = 57 \pm 3 \, \mathrm{km\, s^{-1}}$.

**1.d.ii.** Constraining $\sigma_{RR} : \sigma_{\phi\phi} : \sigma_{zz}$ in the ratio of solar neighbourhood disk dispersions, $39 : 23 : 20$, gives:
(N) $\Omega_b = 27 \pm 3 \, \mathrm{km\, s^{-1} kpc^{-1}}$ and $\sigma_{RR} = 75 \pm 2 \, \mathrm{km\, s^{-1}}$,
(K) $\Omega_b = 26 \pm 2 \, \mathrm{km\, s^{-1} kpc^{-1}}$ and $\sigma_{RR} = 76 \pm 2 \, \mathrm{km\, s^{-1}}$,
(L) $\Omega_b = 24 \pm 2 \, \mathrm{km\, s^{-1} kpc^{-1}}$ and $\sigma_{RR} = 76 \pm 4 \, \mathrm{km\, s^{-1}}$.
All three fits can be rejected at the $2\sigma$ level.

**1.d.iii.** Next we try constraining $\sigma_{RR} : \sigma_{\phi\phi} : \sigma_{zz}$ in the ratio of solar neighbourhood halo dispersions $131 : 102 : 89$. This gives:
(N) $\Omega_b = 26 \pm 2 \, \mathrm{km\, s^{-1} kpc^{-1}}$ and $\sigma_{RR} = 67 \pm 3 \, \mathrm{km\, s^{-1}}$,
This fit can be rejected (just) at the $2\sigma$ level.
(K) $\Omega_b = 26 \pm 2 \, \mathrm{km\, s^{-1} kpc^{-1}}$ and $\sigma_{RR} = 67 \pm 2 \, \mathrm{km\, s^{-1}}$,
This fit can be rejected (just) at the $2\sigma$ level.
(L) $\Omega_b = 24 \pm 2 \, \mathrm{km\, s^{-1} kpc^{-1}}$ and $\sigma_{RR} = 67 \pm 2 \, \mathrm{km\, s^{-1}}$.

It is clear from the above that the assumption of cylindrical rotation, does not significantly affect the fitted rotation and velocity dispersion values listed in sub-paragraph (1.c) above. Again the velocity dispersion model which has the bulge velocity ellipsoid in the shape of the solar neighbourhood disk dispersions is not consistent with the data (at the $2\sigma$ level). The probabilities of the fits are summarised in Table 1. Thus, our data are (just) consistent with a model in which the bulge velocity dispersion is either isotropic or as anisotropic as the solar neighbourhood halo, and where the bulge is a rigid, cylindrical rotator.

### 1.e.  $\sigma_{RR}$ exponential, linear rotation

Next we ask the question: assuming that the bulge is a linear rotator $\overline{v_\phi}(R) = \Omega_b R$, can the velocity dispersion be fit by an exponential function $\sigma_{RR}^2 = \sigma_{RR}^2[R = 0, z = 1.6\,\text{kpc}]\exp(-R/R_s)$? (The scale length $R_s$ is a free parameter in the fit). This behavior of the velocity dispersion with $R$ is observed in thin disk stars (see *e.g.* Gilmore, Wyse & Kuijken 1989). So as to eliminate most of the dependence of $\sigma$ on $z$ in the fits, we consider only the four fields at $b = -12°$.

**1.e.i.** We constrain $\sigma$ such that at a point, $\sigma_{RR} = \sigma_{\phi\phi} = \sigma_{zz}$. The resulting fits have:
(N) $\Omega_b = 24 \pm 2\,\text{km}\,\text{s}^{-1}\text{kpc}^{-1}$, $R_s = 5.2 \pm 2.1\,\text{kpc}$, and
  $\sigma_{RR}[R = 0, z = 1.6\,\text{kpc}] = 71 \pm 10\,\text{km}\,\text{s}^{-1}$,
(K) $\Omega_b = 24 \pm 3\,\text{km}\,\text{s}^{-1}\text{kpc}^{-1}$, $R_s = 4.6 \pm 1.8\,\text{kpc}$, and
  $\sigma_{RR}[R = 0, z = 1.6\,\text{kpc}] = 72 \pm 8\,\text{km}\,\text{s}^{-1}$,
(L) $\Omega_b = 24 \pm 3\,\text{km}\,\text{s}^{-1}\text{kpc}^{-1}$, $R_s = 3.3 \pm 1.3\,\text{kpc}$, and
  $\sigma_{RR}[R = 0, z = 1.6\,\text{kpc}] = 77 \pm 11\,\text{km}\,\text{s}^{-1}$.

**1.e.ii.** We also try $\sigma$ such that at a point, $\sigma_{RR} : \sigma_{\phi\phi} : \sigma_{zz}$ is in the ratio 39 : 23 : 20. Then:
(N) $\Omega_b = 24 \pm 2\,\text{km}\,\text{s}^{-1}\text{kpc}^{-1}$, $R_s = 22.5 \pm 4.9\,\text{kpc}$, and
  $\sigma_{RR}[R = 0, z = 1.6\,\text{kpc}] = 78 \pm 9\,\text{km}\,\text{s}^{-1}$,
(K) $\Omega_b = 24 \pm 3\,\text{km}\,\text{s}^{-1}\text{kpc}^{-1}$, $R_s = 18.2 \pm 4.2\,\text{kpc}$, and
  $\sigma_{RR}[R = 0, z = 1.6\,\text{kpc}] = 79 \pm 8\,\text{km}\,\text{s}^{-1}$,
(L) $\Omega_b = 24 \pm 2\,\text{km}\,\text{s}^{-1}\text{kpc}^{-1}$, $R_s = 8.6 \pm 2.7\,\text{kpc}$, and
  $\sigma_{RR}[R = 0, z = 1.6\,\text{kpc}] = 82 \pm 7\,\text{km}\,\text{s}^{-1}$.
This fit can be rejected at the $2\sigma$ level.

**1.e.iii.** We constrain $\sigma$ such that at a point, $\sigma_{RR} : \sigma_{\phi\phi} : \sigma_{zz}$ is in the ratio 131 : 102 : 89. Then:
(N) $\Omega_b = 24 \pm 2\,\text{km}\,\text{s}^{-1}\text{kpc}^{-1}$, $R_s = 8.2 \pm 3.2\,\text{kpc}$, and
  $\sigma_{RR}[R = 0, z = 1.6\,\text{kpc}] = 76 \pm 5\,\text{km}\,\text{s}^{-1}$,
(K) $\Omega_b = 24 \pm 3\,\text{km}\,\text{s}^{-1}\text{kpc}^{-1}$, $R_s = 7.0 \pm 2.3\,\text{kpc}$, and
  $\sigma_{RR}[R = 0, z = 1.6\,\text{kpc}] = 77 \pm 10\,\text{km}\,\text{s}^{-1}$,
(L) $\Omega_b = 24 \pm 3\,\text{km}\,\text{s}^{-1}\text{kpc}^{-1}$, $R_s = 4.3 \pm 1.5\,\text{kpc}$, and
  $\sigma_{RR}[R = 0, z = 1.6\,\text{kpc}] = 83 \pm 8\,\text{km}\,\text{s}^{-1}$.

It is interesting to note that, especially in the isotropic velocity dispersion situation, the fitted radial scale length $R_s$ is broadly similar to that of the thin disk and that assumed for the thick disk. [Recall that simple dynamical models for the disk suggest that the scale length for variation of the stellar velocity dispersion is just twice that for variation of the spatial density.] Thus with the assumption that $\sigma_{RR}(R)$ is an exponential function, it is possible to fit the bulge with velocity dispersion models where the velocity ellipsoid is as anisotropic as that of the solar neighbourhood disk.

In summary, our data set implies a well constrained linear rotation curve in the bulge, but it allows many bulge velocity dispersion relations; the shape of the velocity ellipsoid can be isotropic or as flattened as the solar neighbourhood disk, and there may, or may not be, radial variations in the velocity dispersion. This is discussed further in section 11.

### 2. Can the outer bulge be triaxial?

Though many studies have found that the bulge is non-axisymmetric (*cf* section 1), it should be stressed that the data we have obtained can be adequately described by simple axisymmetric models. Perhaps when the velocities of many more 'outer bulge' K giants are known, it may be necessary to resort to a less simple explanation. Nevertheless, we investigate whether our data are compatible with non-axisymmetric models.

Published non-axisymmetric bulge models vary significantly according to the bulge tracer they are fit to, and so are not well established. To reduce the large numbers of free parameters, we assume the the bulge has a triaxial form similar to that found by Binney et al.(1991) (which was fit to the central Galactic bar): the axial ratios are $b/a = c/a = 0.75$ and the bar is viewed at an angle of $16°$ to its major axis. We adopt the density profile used by Binney et al.(1991): $\nu(s) = \nu_0(s/s_0)^{-3.5}$, where $s^2 = x^2 + (y/0.75)^2 + (z/0.75)^2$ and $s_0 = 1200\,\text{pc}$. We further assume that the closed '$x_1$' orbits (Contopoulos & Mertzanides 1977) follow approximately the isodensity surface, and that the velocity ellipsoid is aligned with it such that one diagonalising axis ($u$) points out normal to the isodensity surface, another diagonalising axis ($v$) points along the isodensity surface in a plane parallel to the Galactic plane, while the third diagonalising axis ($w$) is perpendicular to the other two. We try models where bulge rotation is a linear function of Galactocentric radius $R$ (independent of $z$ and bulge geometry).

**2.a.** With the above assumptions we can reject at the $4\sigma$ level the most likely fit to a model that has $v_\phi = \Omega_b R$, $\sigma_{uu} = \sigma_{vv} = \sigma_{ww}$ and $\sigma_{uu}(s) = \text{constant}$.

**2.b.** Assuming also that $v_\phi = \Omega_b R$, $\sigma_{uu}(s) = \text{constant}$, $\sigma_{vv}(s) = \text{constant}$ and $\sigma_{ww}(s) = \text{constant}$, we find:
(N) $\Omega_b = 25.4 \pm 4.1\,\text{km}\,\text{s}^{-1}\text{kpc}^{-1}$,
  $\sigma_{uu} = 87 \pm 9\,\text{km}\,\text{s}^{-1}$, $\sigma_{vv} = 30 \pm 6\,\text{km}\,\text{s}^{-1}$ and $\sigma_{ww} = 50 \pm 11\,\text{km}\,\text{s}^{-1}$,
(K) $\Omega_b = 25.0 \pm 3.9\,\text{km}\,\text{s}^{-1}\text{kpc}^{-1}$,
  $\sigma_{uu} = 87 \pm 7\,\text{km}\,\text{s}^{-1}$, $\sigma_{vv} = 28 \pm 6\,\text{km}\,\text{s}^{-1}$ and $\sigma_{ww} = 50 \pm 12\,\text{km}\,\text{s}^{-1}$,
(L) $\Omega_b = 24.8 \pm 3.5\,\text{km}\,\text{s}^{-1}\text{kpc}^{-1}$,
  $\sigma_{uu} = 81\,\text{km}\,\text{s}^{-1}14$, $\sigma_{vv} = 24 \pm 7\,\text{km}\,\text{s}^{-1}$ and $\sigma_{ww} = 84 \pm 48\,\text{km}\,\text{s}^{-1}$.

These fits require a very non-isotropic velocity ellipsoid. This would imply a significant change in the shape of the velocity ellipsoid from our innermost field to Baade's window, where the velocity dispersion is observed to be isotropic (Spaenhauer et al.1992). This calls the physical plausibility of this fit into question, unless the 'nuclear bulge' is unrelated to the 'outer bulge'.

**2.c.** We can reject, again at the $4\sigma$ level, the most likely fit to a model that has $v_\phi = \Omega_b R$, $\sigma_{uu} = \sigma_{vv} = \sigma_{ww}$ and $\sigma_{uu}^2(s) = \sigma_{uu}^2[s = 0]\exp(-s/2s_s)$ (where $s_s$ is a variable).

**2.d.** An isotropic velocity ellipsoid can be fit to our data, if we free the orientation $\alpha$ of the prolate bulge:
(N) $\Omega_b = 24 \pm 4.2\,\text{km}\,\text{s}^{-1}\text{kpc}^{-1}$,
  $\sigma = 56 \pm 7\,\text{km}\,\text{s}^{-1}$ and $\alpha = 29° \pm 6°$,
(K) $\Omega_b = 24 \pm 2.7\,\text{km}\,\text{s}^{-1}\text{kpc}^{-1}$,
  $\sigma = 54 \pm 5\,\text{km}\,\text{s}^{-1}$ and $\alpha = 26° \pm 5°$,

(L) $\Omega_b = 25.2 \pm 3.2 \, \mathrm{km \, s^{-1} kpc^{-1}}$,
$\sigma = 60 \pm 5 \, \mathrm{km \, s^{-1}}$ and $\alpha = 27° \pm 4°$.

Though $\alpha = 27° \pm 5°$ is at odds with the central bar model of Binney et al.(1991), it is consistent with the triaxial HI model of Blitz & Spergel (1991). Thus, with several simplifying assumptions, it is possible to fit our data to bulge models with non-axisymmetric geometry.

The last example above demonstrates the problem of fitting non-axisymmetric models to our data: the constraints on the geometry of the bulge from extant studies allow too great a range of models.

## 8 SPECIFIC ANGULAR MOMENTUM DISTRIBUTION OF THE BULGE

The distribution of specific angular momentum is a primary attribute of a Galactic component (see Wyse & Gilmore 1992). To obtain this distribution one clearly needs the rotation and density distribution through the component. We established in the previous section that the rotation in the bulge from $700 \, \mathrm{pc} \lesssim R \lesssim 3500 \, \mathrm{pc}$ at $b = -12°$ is very well constrained. Almost independent of the model assumed for the velocity dispersion, the bulge rotates at $\Omega_b = 25 \, \mathrm{km \, s^{-1} kpc^{-1}}$. Kent (1992) gives a spatial density model of the bulge which is fit to $2.4 \mu m$ emission for the outer bulge regions and to the data of Allen et al.(1983) and Becklin & Neugebauer (1968) for the central regions. The model has luminosity density $\nu$, such that:

$\nu = 1.04 \times 10^6 \, (s/0.482)^{-1.85} \, \mathrm{L_\odot pc^{-3}}$

for $s < 938 \, \mathrm{pc}$ and

$\nu = K_0(s/667) \, \mathrm{L_\odot pc^{-3}}$

for $s < 938 \, \mathrm{pc}$. In this fit $s$ is an oblate cylindrical coordinate such $(s^4 = R^4 + (z/0.61)^4)$ and $K_0$ is a modified Bessel function.

This density distribution is integrated through the Galaxy, to give the mass $M$ out to radius $R$. From $M(R)$ one can directly obtain the distribution of specific angular momentum $M(h)$ by substituting for the angular momentum $h = \Omega_b R^2$. The distribution of specific angular momentum of the bulge is displayed in Figure 8, where we also show the specific angular momentum distribution of the halo, disk and thick disk, as given by Wyse & Gilmore (1992). The bulge, under the above assumptions, has an even more dissipated specific angular momentum distribution than found by Wyse & Gilmore (1992). This is of course consistent with the hypothesis that the bulge is the dissipated core of the halo, but is not consistent with the hypothesis that the bulge is closely related, in an evolutionary sense, to the thick disk or the thin disk.

## 9 THE BULGE METALLICITY DISTRIBUTION FUNCTION

We show a representative colour-magnitude (CM) diagram of the fields studied. Figure 9 is the CM diagram in $(B_J - R)$ and R as found for the $(\ell = 5°, b = -12°)$ field. The 'star' symbols represent the stars observed spectroscopically. In this field, the distance modulus to the tangent point to the line of sight is $(m - M) = 14.5$, and the reddening is $E_{B-V} = 0.14$. By superimposing CM relations for four globular clusters of different metallicities (shifted to the tangent point), we illustrate the metallicity range of stars that have been observed near the bulge. (The colour-magnitude system $(B_J - R)$, R was converted into $(B - V)$, V using the colour equations presented in Paper I). The clusters are from left to right: M92, M3, M5, 47 Tucanae and NGC 188. At the tangent point in this field we therefore expect to find few halo stars significantly more metal poor than M3 ([Fe/H] = $-1.66$ – Djorgovski 1993), but more metal rich populations should easily be detected, if present.

In a high-resolution spectroscopic study of bulge K-giants, McWilliam & Rich (1994) find that Mg and Ti in their stars are more abundant by $\approx 0.3$ dex than in the Sun over the range of [Fe/H] they observed, while Ca and Si are present in similar ratios to those of disk giants. If that deduction is correct, we cannot simply infer [Fe/H] from [Mg/H], as measured by the Mg index defined in Paper I, which was calibrated against local K-giants. Further data would be necessary to establish the bulge [Mg/Fe] gradient. When the correctness of the bulge element ratio results has been established, a correction will need to be applied to the abundances derived here.

With the above health warning, we show the relationship between metallicity and radial velocity for objects classified as K giants in Figure 10. Recall that reliable calibration of the photometry is not available for the fields $(\ell = -15, b = -12)$ and $(\ell = 5, b = -20)$, so that for these fields [Fe/H] is really only a ranking in order of relative abundance. Furthermore, we show in IGI that the low velocity dispersion feature with mean velocity $\approx 170 \, \mathrm{km \, s^{-1}}$ seen in Figure 10d, Figure 10e and Figure 10f is the Sagittarius dwarf, an object that is not part of the Milky Way, at least, just yet. We therefore can confidently analyse the abundance – radial velocity data of Figure 10a, Figure 10c, Figure 10d and Figure 10e , and for the latter two fields we select only those stars with Galactocentric radial velocity $v < 130 \, \mathrm{km \, s^{-1}}$: we proceed to do this in the remainder of this section.

Recall that $\approx 90\%$ of objects in our sample classified as K dwarfs have formally-calculated [Fe/H] > 0.0. The technique for determining these values was calibrated on K giants only, so for K dwarfs the calculated 'metallicity' values will be very different from the actual metallicity values. In most fields, the expected number of dwarf stars will be $\approx 20\%$, while (mainly) photometric errors will induce $\approx 25\%$ misclassification. Given that $\approx 20\%$ of objects classified as K giants have [Fe/H] > 0.0, we should expect $\approx 25\%$ K dwarf contamination in the K giant sample for stars with calculated [Fe/H] > 0.0. Thus, the observed metallicity distributions are unreliable for [Fe/H] > 0.0.

The velocity dispersion in the four fields at $(\ell = -25°, b = -12°)$, $(\ell = -5°, b = -12°)$, $(\ell = 5°, b = -12°)$ and $(\ell = 5°, b = -15°)$ decreases with increasing metallicity. We divide the stars into three bins; these are defined as:

i) bin-1. [Fe/H] < $-1.68$ — that is, a region containing stars more metal poor than the mean halo abundance (Laird et al.1988)

ii) bin-2. $1.68 <$ [Fe/H] $< 0.0$ — the region between the mean halo abundance and the abundance where we expect dwarf and distant disk giant contamination to be-

**Table 1:** KS test probabilities for the bulge functional fits for all the models described in the text.

| Test | $(-25°, -12°)$ | $(-15°, -12°)$ | $(-5°, -12°)$ | $(5°, -12°)$ | $(5°, -15°)$ | $(5°, -20°)$ |
|---|---|---|---|---|---|---|
| 1.a.i    | 0.52 | 0.69 | 0.47 | —    | —    | —    |
| 1.a.ii   | 0.45 | 0.53 | 0.58 | —    | —    | —    |
| 1.a.iii  | 0.46 | 0.41 | 0.66 | —    | —    | —    |
| 1.b.i(N)   | — | — | — | 0.43 | 0.11 | 0.96 |
| 1.b.i(K)   | — | — | — | 0.45 | 0.13 | 0.97 |
| 1.b.i(L)   | — | — | — | 0.33 | 0.16 | 0.96 |
| 1.b.ii(N)  | — | — | — | 0.43 | 0.14 | 0.96 |
| 1.b.ii(K)  | — | — | — | 0.12 | 0.15 | 0.14 |
| 1.b.ii(L)  | — | — | — | 0.55 | 0.21 | 0.17 |
| 1.b.iii(N) | — | — | — | 0.38 | 0.14 | 0.97 |
| 1.b.iii(K) | — | — | — | 0.12 | 0.15 | 0.97 |
| 1.b.iii(L) | — | — | — | 0.23 | 0.11 | 0.97 |
| 1.c.i(N)   | 0.13 | 0.28 | 0.10 | 0.29 | — | — |
| 1.c.i(K)   | 0.11 | 0.25 | 0.07 | 0.35 | — | — |
| 1.c.i(L)   | 0.08 | 0.11 | 0.03 | 0.40 | — | — |
| 1.c.ii(N)  | 0.28 | 0.41 | 0.07 | 0.43 | — | — |
| 1.c.ii(K)  | 0.26 | 0.37 | 0.08 | 0.22 | — | — |
| 1.c.ii(L)  | 0.17 | 0.21 | 0.04 | 0.27 | — | — |
| 1.c.iii(N) | 0.19 | 0.34 | 0.05 | 0.22 | — | — |
| 1.c.iii(K) | 0.16 | 0.31 | 0.07 | 0.27 | — | — |
| 1.c.iii(L) | 0.10 | 0.15 | 0.18 | 0.33 | — | — |
| 1.d.i(N)   | 0.43 | 0.19 | 0.28 | 0.42 | 0.15 | 0.12 |
| 1.d.i(K)   | 0.20 | 0.27 | 0.09 | 0.32 | 0.08 | 0.09 |
| 1.d.i(L)   | 0.10 | 0.12 | 0.15 | 0.37 | 0.10 | 0.11 |
| 1.d.ii(N)  | 0.46 | 0.41 | 0.06 | 0.15 | 0.08 | 0.07 |
| 1.d.ii(K)  | 0.41 | 0.38 | 0.12 | 0.20 | 0.09 | 0.05 |
| 1.d.ii(L)  | 0.27 | 0.21 | 0.07 | 0.24 | 0.11 | 0.04 |
| 1.d.iii(N) | 0.32 | 0.35 | 0.10 | 0.20 | 0.07 | 0.15 |
| 1.d.iii(K) | 0.28 | 0.32 | 0.13 | 0.26 | 0.10 | 0.08 |
| 1.d.iii(L) | 0.16 | 0.16 | 0.18 | 0.30 | 0.11 | 0.19 |
| 1.e.i(N)   | 0.97 | 0.13 | 0.18 | 0.49 | — | — |
| 1.e.i(K)   | 0.99 | 0.22 | 0.15 | 0.53 | — | — |
| 1.e.i(L)   | 0.97 | 0.34 | 0.06 | 0.67 | — | — |
| 1.e.ii(N)  | 0.71 | 0.12 | 0.20 | 0.71 | — | — |
| 1.e.ii(K)  | 0.92 | 0.25 | 0.17 | 0.45 | — | — |
| 1.e.ii(L)  | 0.96 | 0.29 | 0.11 | 0.37 | — | — |
| 1.e.iii(N) | 0.89 | 0.11 | 0.20 | 0.43 | — | — |
| 1.e.iii(K) | 0.99 | 0.23 | 0.16 | 0.47 | — | — |
| 1.e.iii(L) | 0.98 | 0.29 | 0.18 | 0.42 | — | — |
| 2.b(N) | 0.87 | 0.36 | 0.16 | 0.66 | 0.34 | 0.33 |
| 2.b(K) | 0.65 | 0.41 | 0.19 | 0.78 | 0.35 | 0.29 |
| 2.b(L) | 0.51 | 0.20 | 0.11 | 0.85 | 0.38 | 0.31 |
| 2.d(N) | 0.87 | 0.36 | 0.16 | 0.66 | 0.34 | 0.33 |
| 2.d(K) | 0.65 | 0.41 | 0.19 | 0.78 | 0.35 | 0.29 |
| 2.d(L) | 0.42 | 0.15 | 0.18 | 0.71 | 0.23 | 0.49 |

**Figure 8.** The specific angular momentum distributions for the four major Galactic components. The solid curve is the distribution of the bulge component, as given by the rotational velocity found in this work together with a density model of Kent (1992). The other curves are taken from Wyse & Gilmore (1992); the dashed-dotted curve represents the halo, the dotted curve represents the thick disk and the dashed curve represents the disk.

**Figure 9.** ($B_J$ – R, R) colour-magnitude diagram of stars measured in a $1° \times 1°$ region of sky in the ($\ell = 5°, b = -12°$) field. The 'star' symbols represent the stars observed spectroscopically. The four colour-magnitude relations superimposed are, from left to right: M92, M3, M5, 47 Tucanae (from Sandage 1982) and M67 (from Chiu 1980), all shifted to the distance of the tangent point in this line of sight.

come significant

iii) bin-3. [Fe/H] > 0.0 — where we expect significant dwarf and distant disk giant contamination

The velocity dispersions in Figure 10a are found to be $85.7\,\text{km s}^{-1}$, $59.5\,\text{km s}^{-1}$ and $46.4\,\text{km s}^{-1}$ in the first, second and third bins respectively, while in Figure 10c the velocity dispersions are $112.3\,\text{km s}^{-1}$, $70.9\,\text{km s}^{-1}$ and $58.3\,\text{km s}^{-1}$ in the first, second and third bins respectively. We do not calculate similar quantities in Figure 10d and Figure 10e, because of the contamination in those fields.

We model the metallicity distribution of the halo as a gaussian of intrinsic width $\sigma = 0.2$dex centred at $\overline{\text{Fe/H}} = -1.68$ (Laird et al.1988), and we assume that the shape of the thick disk metallicity distribution towards the South Galactic Pole (Gilmore, Wyse & Jones 1995) is the same as that in the fields studied here. That is, we assume that there are no radial abundance gradients in either the halo or the thick disk. So we pose the question: is it possible to fit the observed metallicity distributions with a composite model that consists of the sum of the above halo and thick disk models together with a 'simple model' of chemical evolution to represent the bulge?

We first need to know the relative numbers of halo and thick disk stars in our sample: in table 2, we show the number of halo and thick disk stars predicted by the galaxy model in the four fields investigated.

The simple, or closed-box, model of Galactic chemical evolution is discussed in standard texts, and is discussed by Rich (1990) in the context of the Milky Way bulge. The probability density function $dN/d$[Fe/H] it predicts is given

**Figure 10.** The relationship between velocity and metallicity for stars classified as giants. The fields are: (a) $\ell = -25°$, $b = -12°$, (b) $\ell = -15°$, $b = -12°$, (c) $\ell = -5°$, $b = -12°$, (d) $\ell = 5°$, $b = -12°$, (e) $\ell = 5°$, $b = -15°$ and (f) $\ell = 5°$, $b = -20°$.

by

$$\frac{dN}{d[\text{Fe/H}]} = \frac{z_\odot \ln 10}{y} \exp(-\frac{z_\odot}{y} 10^{[\text{Fe/H}]}) 10^{[\text{Fe/H}]}, \qquad (6)$$

where $y$ is the yield and $z_\odot$ is the solar metal abundance. In an evolved population which has turned its gas into stars, the mean abundance equals the yield, *i.e.* $\overline{z} = y$.

Since we are not confident of the observed distributions in the region [Fe/H] > 0.0, we will truncate both the model and the data at [Fe/H] = 0.0. We stress that this cutoff will have the effect of altering the yield of the resulting fits. However, it is useful to have a simple parameterisation of the data, and this specific limitation in application is relatively minor compared to the rather extreme nature of some of the physical assumptions underlying the simple model.

To simulate measuring errors in the above composite model, we convolve the empirical thick disk distribution of Gilmore, Wyse & Jones (1995) with a gaussian of $\sigma = 0.25$ (the difference of the measuring errors between their and our [Fe/H] measurements), while the theoretical bulge and halo distributions are convolved with our metallicity error distribution (*cf* Paper I). The only free parameter is the yield of the 'simple model', which we find by maximising the likelihood of the fit. The results are shown in Figure 11; all these fits are acceptable at better than the $1.5\sigma$ level. The yield $y$ in units of $z_\odot$ is found to be $0.69 \pm 0.20$, $0.43 \pm 0.11$, $0.53 \pm 0.06$ and $0.46 \pm 0.09$ in Figure 11a, Figure 11b, Figure 11c and Figure 11d respectively. The reader can use the agreement or otherwise between these derived yields as an indication of the ability of the model to represent the data adequately.

**Table 2.** The predicted number of halo and thick disk stars with [Fe/H] < 0.0 according to the Galaxy model. In the $(5°, -12°)$ and $(5°, -15°)$ fields, we have also only selected stars with Galactocentric radial velocity $v < 130\,\mathrm{km\,s^{-1}}$.

| Component | $(-25°, -12°)$ | $(-5°, -12°)$ | $(5°, -12°)$ | $(5°, -15°)$ |
|---|---|---|---|---|
| Total | 158 | 206 | 219 | 66 |
| Halo | 3.5 | 4.3 | 4.8 | 1.2 |
| Thick disk | 32.2 | 30.3 | 32.2 | 5.6 |

The kinematic information can now be used to check for self-consistency in this model. As usual, we will not use the fields at $(\ell = 5°, b = -12°)$ and $(\ell = 5°, b = -15°)$, because of the contamination from the dwarf galaxy. Note that according to the adopted thick disk metallicity distribution, less than 0.5% of stars in bin-1 are thick disk stars, while thin disk giants comprise less than 5% of the stars in bin-2 and contribute negligibly to bin-1. We then fit the kinematics of bulge stars in bins 1 and 2 separately using our Galaxy model; standard kinematic and space density distributions are assumed for the halo and thick disk (with normalisations in bins 1 and 2 as detailed above), while the bulge component is modeled as an oblate rotator with isotropic velocity dispersion throughout. Descriptions of these distributions may be found in Gilmore, King & van der Kruit (1990) and in Gilmore, Wyse & Kuijken (1989). The bulge is assumed to rotate rigidly at a rate $\Omega = 25\,\mathrm{km\,s^{-1}\,kpc^{-1}}$ (as derived in §7.3), which leaves the bulge dispersion as the only free parameter in the fit.

A maximum likelihood fit to the data in the $(\ell = -25, b = -12)$ field gives $\sigma = 82 \pm 46$ in bin-1 and $\sigma = 69 \pm 7$ in bin-2, while a fit to the data in the $(\ell = -5, b = -12)$ field gives $\sigma = 139 \pm 34$ in bin-1 and $\sigma = 53 \pm 6$ in bin-2. We therefore conclude that either there are significantly more halo stars in the field at $(\ell = -5, b = -12)$ than the standard halo model predicts (so as to increase the ratio of halo to bulge stars in bin-1), or there is a significant dependence of velocity dispersion on metallicity in the bulge.

Alternatively, if we demand that there is no change in bulge dispersion with metallicity in the $(\ell = -5, b = -12)$ field, then there must be approximately 25 halo stars in the sample (a factor of $\approx 5$ more than the standard model), and the bulge dispersion becomes $99 \pm 30\,\mathrm{km\,s^{-1}}$, consistent with $53 \pm 6\,\mathrm{km\,s^{-1}}$. If this assumption is true, we find that the composite metallicity model described above, which has a 'simple model' with yield $y = 0.72 \pm 0.26$ representing the bulge component, is acceptable at better than the $1\sigma$ level; the fit is shown in Figure 12.

How much are the above fits affected by our decision to ignore bin-3 data? We take the numbers of halo and thick disk stars as given in table 2, then fit the composite model to the data in all 3 bins. The resulting yield (in units of $z_\odot$) is found to be $y = 0.76 \pm 0.07$, $y = 0.40 \pm 0.05$, $y = 0.64 \pm 0.08$ and $y = 0.74 \pm 0.10$ in the field at $(\ell = -25°, b = -12°)$, $(\ell = -5°, b = -12°)$, $(\ell = 5°, b = -12°)$ and $(\ell = 5°, b = -15°)$ respectively (the distributions are acceptable at better than the $1.5\sigma$ level). Comparison with the results above shows that imposing the abundance cut marginally reduces the derived yield, as expected.

In summary, the yield found in each field is dependent on the imposed metallicity cut and slightly dependent on the assumed normalisation of thick disk and halo stars. Nevertheless, it appears that the bulge component can be adequately well described (for [Fe/H] < 0.0) by the abundance distribution predicted by the 'simple' or closed box model of chemical evolution. From fitting this simple model we obtain the yield of a simple-model fit to the K giant population in the neighbourhood of the tangent point in each field: this is consistent with no metallicity gradient through the region in which we have data: *i.e.* $0.7\,\mathrm{kpc} \lesssim R \lesssim 3.5\,\mathrm{kpc}$. Recall also that recent developments (McWilliam & Rich 1994) have shown that K giants in Baade's window typically have abundance [Fe/H] $\approx -0.25$ — this would imply a zero metallicity gradient throughout almost the whole of the bulge.

### 9.1 A Connection with the Thick Disk?

Since the thick disk metallicity distribution in Figure 11 looks similar to that of the bulge, we will investigate whether we can fit the observed metallicity distributions with a model similar to that discussed above, but whose bulge component has the same metallicity distribution as the thick disk.

We find a good (better than $1.5\sigma$) fit by adopting the Galaxy model halo star prediction of 4 halo stars in the $(\ell = -25°, b = -12°)$ field; this is displayed in Figure 13a. If the model predictions for the number of halo stars in the other fields are correct, then we can reject at the $2\sigma$ level the hypothesis that the bulge has the same abundance distribution as the thick disk in those fields. To obtain agreement at better than the $1.5\sigma$ level, we require 30 halo stars in the $(\ell = -5, b = -12)$ field, 40 halo stars in the $(\ell = 5, b = -12)$ field and 20 halo stars in the $(\ell = 5, b = -15)$ field. Recall that 4 such stars are expected. These fits are shown in Figure 13.

**Figure 11.** The metallicity distributions in (a) the field at $(\ell = -25°, b = -12°)$, in (b) at $(\ell = -5°, b = -12°)$, in (c) at $(\ell = 5°, b = -12°)$ and in (d) at $(\ell = 5°, b = -15°)$. These are compared to the model described in the text, in which the bulge component is represented by a simple 'closed box' chemical evolution model, with a yield fitted independently to each field. In each panel, the dashed line represents the halo component, the dashed-dotted line represents the thick disk, the dotted line represents the bulge and the full line is the sum of all three components.

## 10  COMPARISON TO EXTERNAL GALAXIES

Kormendy & Illingworth (1982; hereafter KI82) compared the importance of rotational support in spiral galaxy bulges to that in elliptical galaxies. The ratio of the energy in ordered motion to that in random motion was estimated from the ratio of the maximum rotational velocity to the central velocity dispersion: $V_m/\sigma_0$. $V_m/\sigma_0$ is approximately equal to the mass-weighted averaged quantities $(\overline{V^2}/\overline{\sigma}^2)^{1/2}$ in isotropic oblate spheroid models (Binney 1980). It is the mean stellar rotation that provides support and flattening in these models. Similarly, Davies et al.(1983a) showed that the kinematics of elliptical galaxies of low absolute luminosity are also well described by such models, while intrinsically bright elliptical galaxies bear most resemblance to models which are supported by random motion (ie pressure) and flattened by anisotropic velocity dispersions.

An interesting feature of the data of KI82 is the lack of spatial overlap between their data for external bulges with extant data for the Milky Way. The only regions of the Milky Way not dominated by the disk where the kinematics of K-giants were well known are the Baade's Window bulge and the subdwarf halo in the solar neighbourhood. Comparable regions in the KI82 edge-on spiral galaxies were either hidden by the galactic disks or were too faint to be detected. KI82 quote the limiting surface brightness in their data as 24 B mag arcsec$^{-2}$, while Morrison (1993) estimates that the surface brightness of the halo, as viewed edge-on from outside the Galaxy at the solar radius, is $\mu_V = 27.7$ V mag arcsec$^{-2}$. The various contributions to the light profile observed by KI82, and the consequences thereof, were discussed in some detail by Shaw & Gilmore (1989). The essen-tial point is that no data were available to compare available spectroscopic studies of external spiral bulges with the bulge of the Milky way. We now have the necessary kinematic data to compare the Milky Way bulge to the spiral galaxy bulges observed by KI82, over a similar range of $R/h_R$ and $z/h_z$, where $h_R$ and $h_z$ are characteristic bulge scale lengths in the $R, z$ directions. We also investigate whether the Milky Way bulge is typical of spiral galaxy bulges. Since the bulge spatial density profile is now reasonably well known (e.g. Kent 1992; Freeman 1993), we also investigate whether the above-mentioned assumption, used by KI82, that $V_m/\sigma_0 \approx \overline{V}_m/\overline{\sigma}$, holds true in our Galaxy.

**Figure 12.** Comparison of the data towards ($\ell = -5°, b = -12°$) with the composite model described in the text. We have assumed that there are 25 halo stars in the sample rather than the 4 expected. The dashed line represents the halo component, the dashed-dotted line represents the thick disk, the dotted line represents the bulge and the full line is the sum of all three components.

In most of the galaxies they studied, KI82 were not able to measure $V_m$ and $\sigma_0$ directly due to problems with disk contamination, but used instead interpolations to estimate those parameters. We reproduce their technique on the Milky Way bulge.

KI82 calculated $V_m$ from those bulge regions where "$V(r)$ reaches a well-defined plateau or maximum at radii dominated by bulge light". This is the case in our ($l = -25°, b = -12°$) field, where the mean line of sight velocity has increased by less than $15\,\text{km s}^{-1}$ over a range of $\approx 1\,\text{kpc}$ from ($l = -15°, b = -12°$). Since the mean rotation in most galaxy bulges decreases with height above the galactic plane, KI82 extrapolated the $V_m$ values down to $V_m(z = 0)$. The extrapolation was performed by using the functional form of $V(z)$ as obtained from two-dimensional velocity maps of elliptical galaxies by Davies and Illingworth (1983b). $V(z)$ from these elliptical galaxies was scaled by the length $z_e$, the value of the effective radius of a Vaucouleurs $r^{1/4}$ law fit to the minor axis profile. We use $z_e = 200\,\text{pc}$ for the Milky Way bulge from Frogel et al.(1990). Where the external galaxy bulge displayed cylindrical rotation, KI82 used $V_m(z = 0) = V_m$. However, whether or not the Milky Way bulge is a cylindrical rotator is not as yet established (cf section 4), so we will treat both cases separately. Cylindrical rotation would then give $V_m(0) = 95\,\text{km s}^{-1}$ at ($\ell = -25°, b = 0°$), while applying the above $V(z)$ correction for the non-cylindrical rotation case, we obtain $V_m(0) = 104$.

KI82 approximate central dispersion $\sigma_0$ to an average $\overline{\sigma}$ out to $\approx r_e/2$, where $r_e$ is the effective radius of a de Vaucouleurs (1959) $r^{1/4}$ law fit. No K giant data are available over the same Galactocentric distance range. However, central Miras $r \approx 350\,\text{pc}$ from the Galactic centre have velocity dispersion $127 \pm 24\,\text{km s}^{-1}$ Feast et al.(1990). This value is compatible with the velocity dispersion of $131 \pm 28\,\text{km s}^{-1}$ of Carbon stars at $r \approx 450\,\text{pc}$ found by Tyson & Rich (1991).

Thus $V_m/\sigma_0 = 0.75$ if the bulge is a cylindrical rotator, or $V_m/\sigma_0 = 0.82$ if it resembles the other model above.

Therefore, if the bulge is similar to the oblate rotating models with isotropic velocity dispersion (Binney 1978), to which KI82 compare their data (see KI82, their figure 3), then we find that the ellipticity of the Milky Way bulge should be $\epsilon \approx 0.4$. This value of bulge ellipticity is very similar to the value $\epsilon = 0.39$ found by Kent et al.(1991).

## 11 CONCLUSIONS

Our analysis of the data discussed in Paper I has shown that the outer regions of the Galactic bulge between $0.7\,\text{kpc} \lesssim R \lesssim 3.5\,\text{kpc}$ and $1.2 \lesssim |z| \lesssim 2.7\,\text{kpc}$ can be adequately described by axisymmetric models with linear rotation $\Omega = 25\,\text{km s}^{-1}\,\text{kpc}^{-1}$. The data also constrain the Galactic disk to have vertical scale height $z_h \lesssim 220\,\text{km s}^{-1}$ in the central $\approx 3.5\,\text{kpc}$. Since the vertical scale height of the Galactic disk in the solar neighbourhood is $z_h = 250\,\text{pc}$ (Kuijken & Gilmore 1989), the Galactic disk appears to be different from pure-disk galaxies, where the disk scale height is independent of galactocentric distance. Alternatively, our selection function for K giants continues to select preferentially in favour of the young disk in preference to the old disk. The data allow several velocity dispersion $\sigma(R)$ models. If $\sigma(R)$ is isothermal, then the bulge velocity ellipsoid can only be about as anisotropic as that of the solar neighbourhood halo. However, if $\sigma(R)$ decreases exponentially with $R$, the bulge velocity ellipsoid can be as anisotropic as that of the solar neighbourhood disk. More data in several more fields will be required to confirm or reject cylindrical rotation and to constrain better the radial dependence of the shape, size and orientation of the velocity ellipsoid. The simplest allowed model which is consistent with this data-set does not require non-axisymmetric geometry.

We find acceptable agreement with the axisymmetric bulge model devised by Kent (1992), which uses a Galactic mass model fit to $2.4\mu m$ luminosity, with the assumption $M/L = 1$, to predict self-consistently the kinematic structure of the bulge. Thus the potential of the inner Galaxy is currently reasonably well established.

We show that the shape of the observed abundance distribution (for [Fe/H] < 0.0, where we believe are data are reliable) can be adequately represented by a model that consists of the sum of the abundance distributions for the halo and thick disk observed in the solar neighbourhood, together with an abundance distribution for the bulge as predicted by the 'simple model' of chemical evolution. The maximum-likelihood fitted yield in each field does not vary significantly over the Galactic regions for which we have data, and is consistent with a zero metallicity gradient in the bulge K giant population from the Baade's Window region to $R \approx 3.5\,\text{kpc}$. Thus the bulge is a well-mixed population.

**Figure 13.** We test the hypothesis that the bulge has a similar metallicity distribution to the empirical thick disk distribution at the Solar Galactocentric distance determined by Gilmore, Wyse & Jones (1995). The metallicity distributions correspond to (a) the field at $(\ell = -25°, b = -12°)$, in (b) to $(\ell = -5°, b = -12°)$ in (c) $(\ell = 5°, b = -12°)$ and in (d) $(\ell = 5°, b = -15°)$. The number of halo stars required to make these fits acceptable at better than the $1.5\sigma$ level, are 3.5, 30, 40, and 5 stars in (a), (b), (c) and (d) respectively, while $\sim 4$ are expected in each field. The dashed line represents the halo component, the dotted line represents the bulge and the full line is the sum of the two components.

The bulge region investigated in this volume covers approximately the same Galactocentric distance range (measured in bulge scale lengths) as did the kinematic study of external spiral galaxy bulges presented by Kormendy & Illingworth (1982). Our data allow the first detailed kinematic comparison of the Milky Way bulge to the population of external spiral galaxy bulges. We find that the Galactic bulge is representative of the population of external spiral galaxy bulges.

With our measured linear rotation curve, and adopting the mass model derived from the observed near-IR surface brightness of the bulge by Kent (1992), we derive the specific angular momentum distribution of the bulge. This distribution is very similar to, though slightly more dissipated than, that of the halo, and so is consistent with the hypothesis that the bulge is the dissipated core of the halo. The corresponding distributions of specific angular momentum for the disks are very different. Thus, our results are not easily consistent with any model in which the bulge is closely related, in an evolutionary sense, to the Galactic disks.

Finally, we note that our study is of what we have termed the *Outer Bulge*. Although this includes most of the volume of the central Galactic bulge, in the sense that this term is applied to external galaxies, most recent studies of the central Galaxy have been more concerned with the very inner parts of the Bulge. This part of the Galaxy is nicely seen in recent IRAS and COBE maps, but is nearly completely obscured from short wavelength, star by star study. This very central, and very compact, structure is the central highest density part of all Galactic structural components.

It is possible that something else also lives there. We are confident that the order of magnitude improvements in spatial resolution and sensitivity which will be provided by ISO at wavelengths appropriate to study the dominant central stellar population will provide substantial improvements in our knowledge of this very central Galactic core. In the interim, we refrain from speculation as to the relationship, if any, between that region and the outer bulge which has been the subject of this investigation.

## ACKNOWLEDGEMENT

We thank Steve Kent for providing us with the toolkit to facilitate comparison of his mass model with our kinematic data.

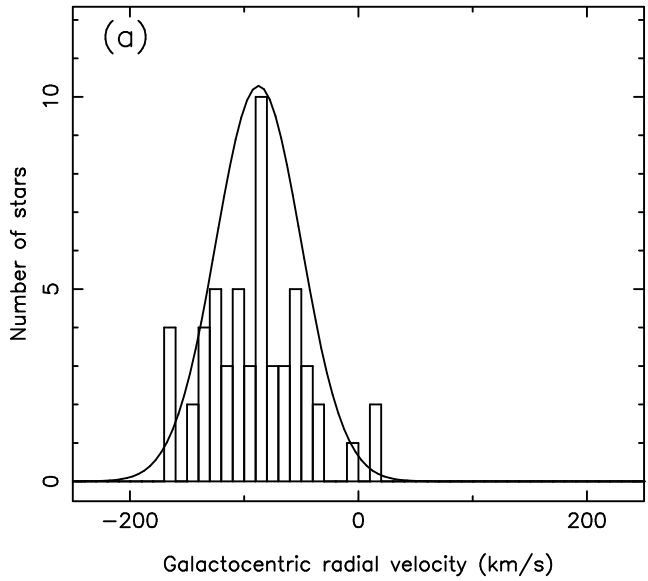
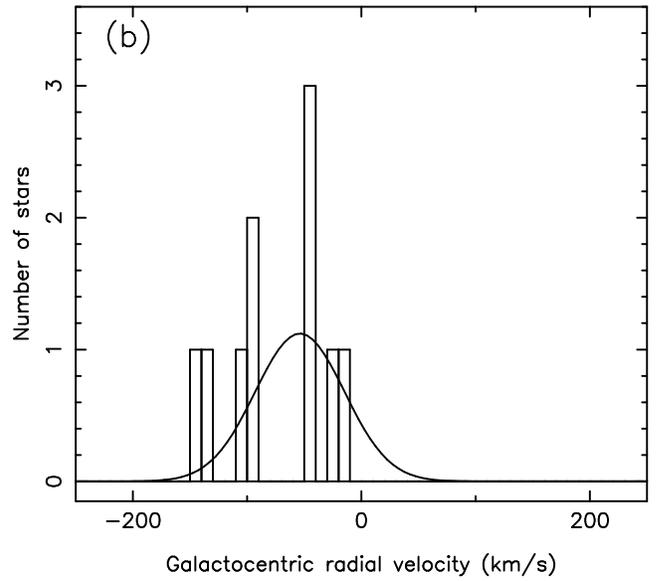
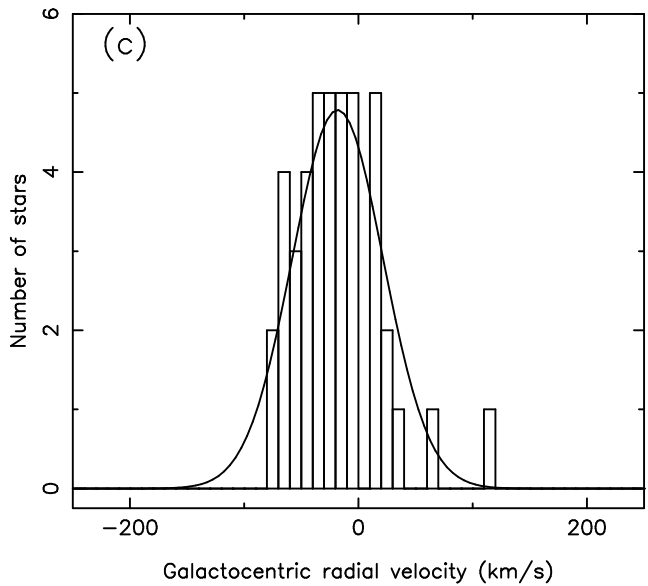
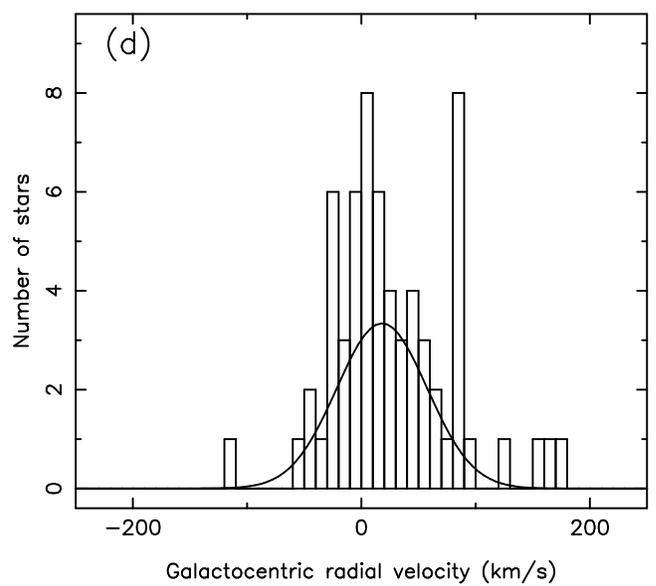
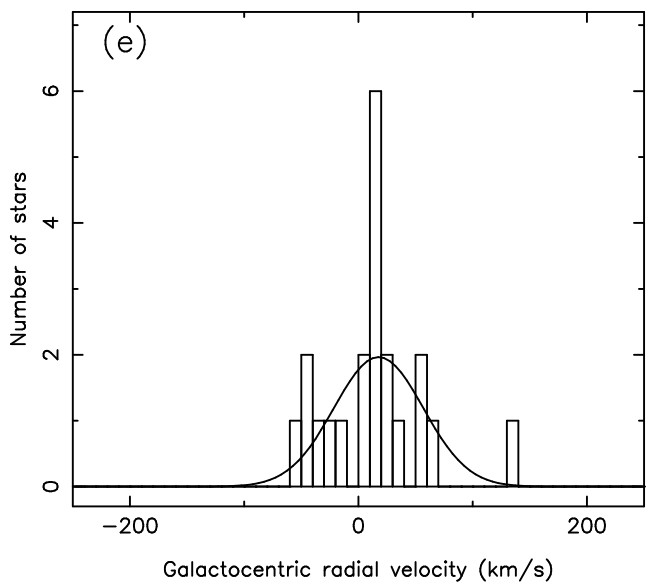
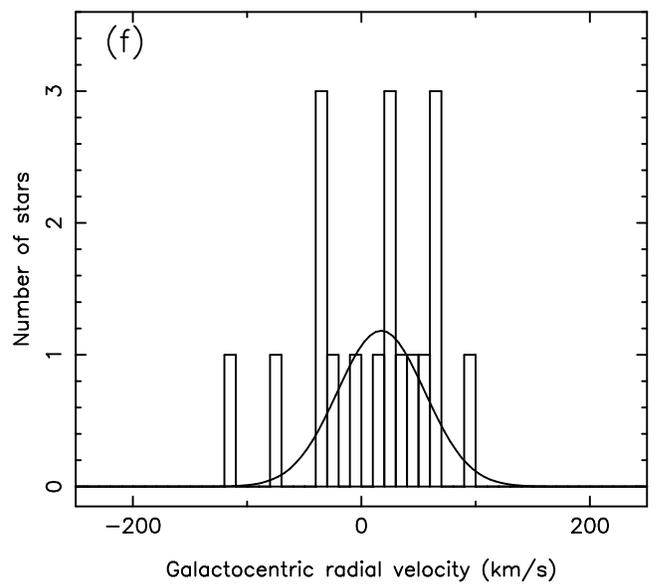

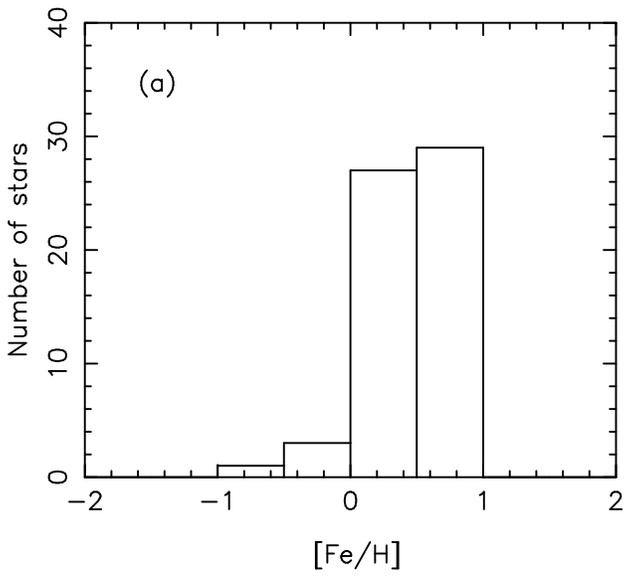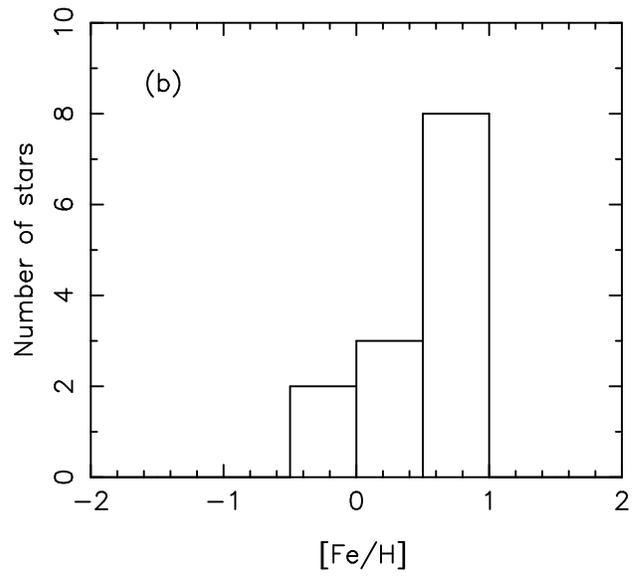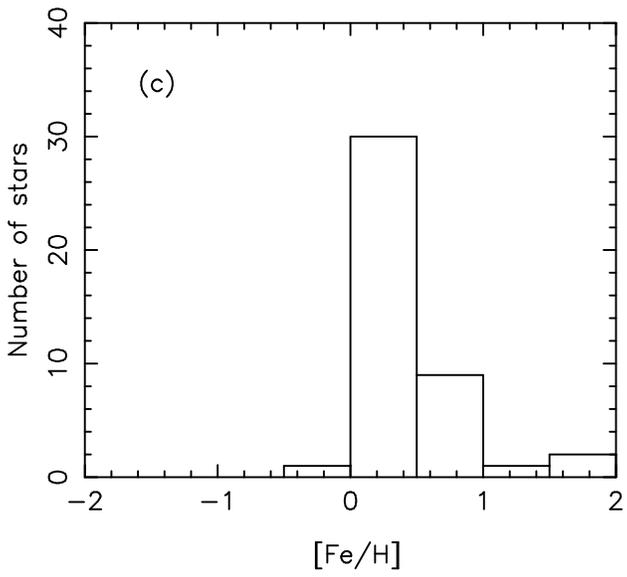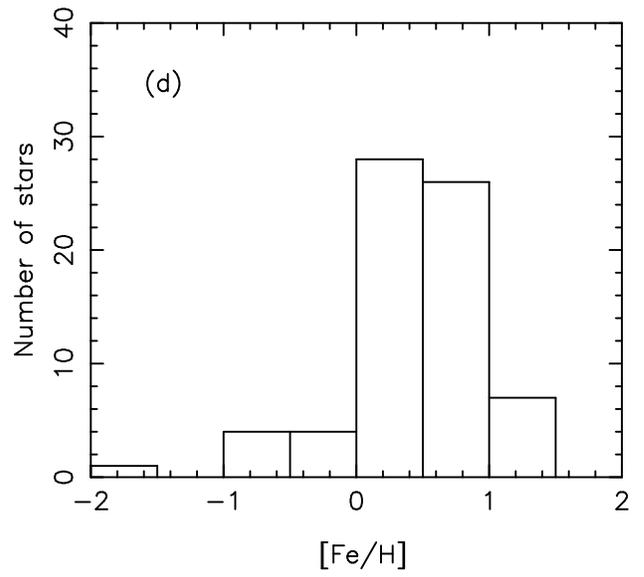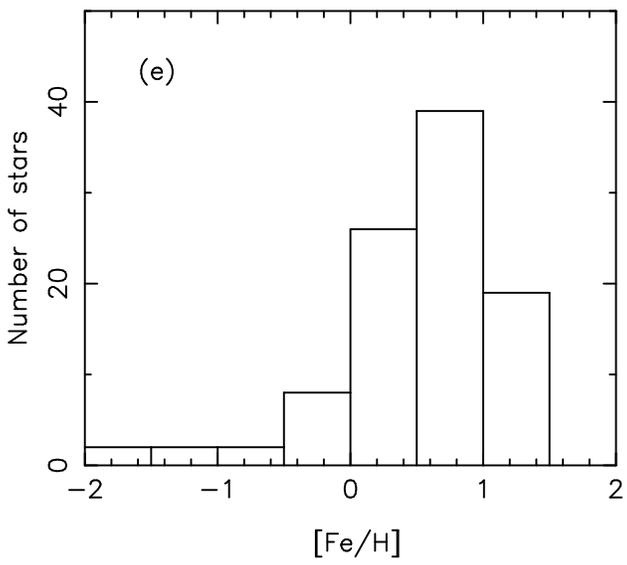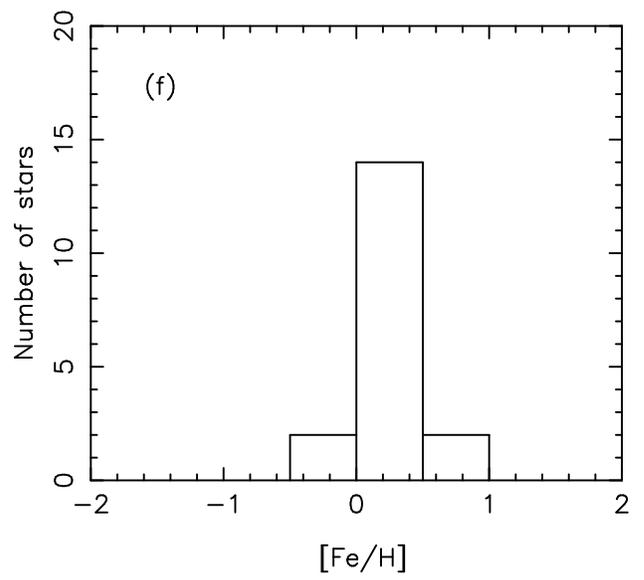

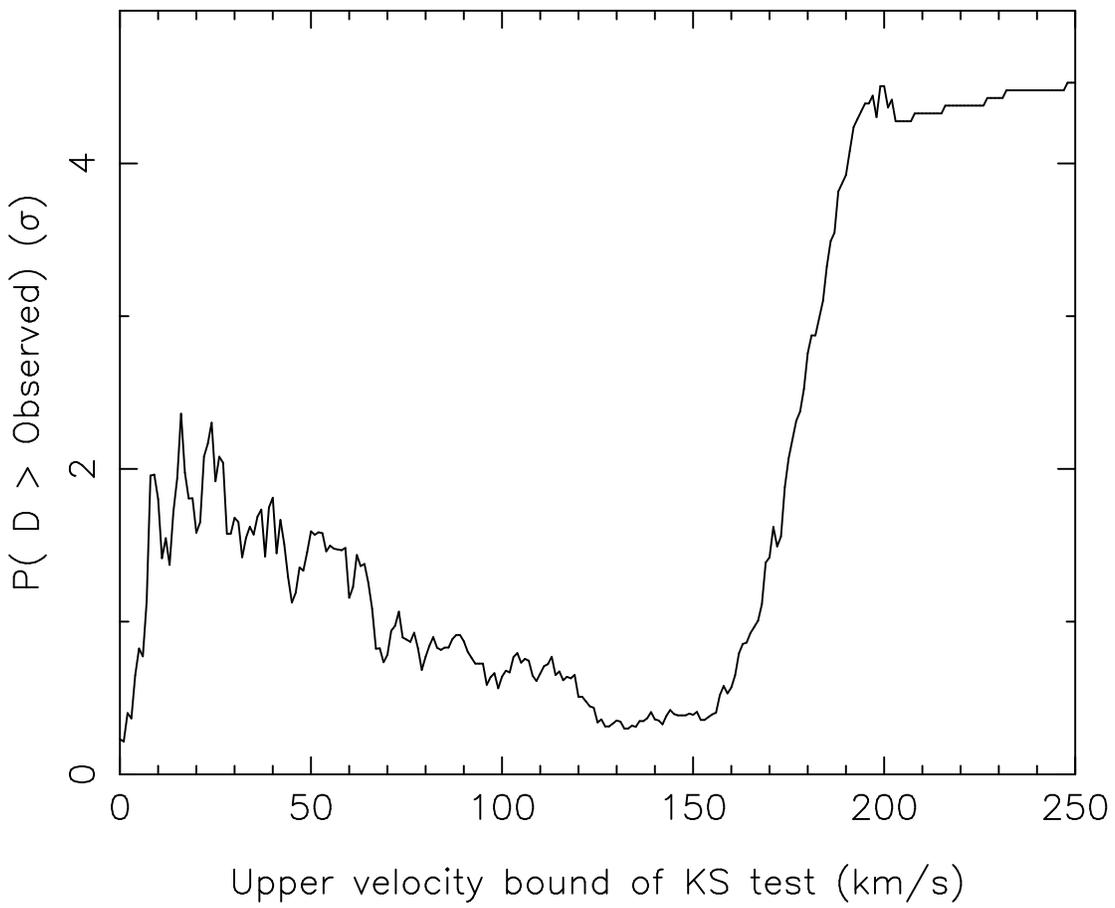

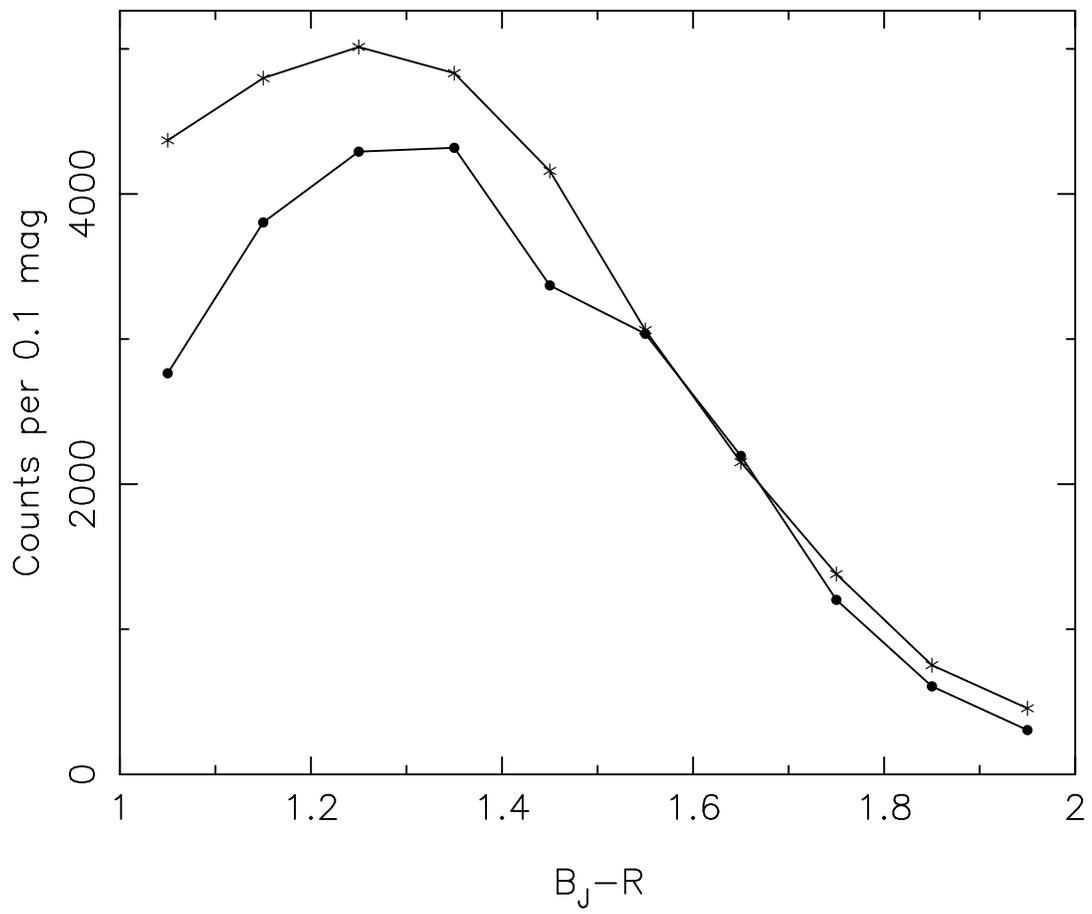

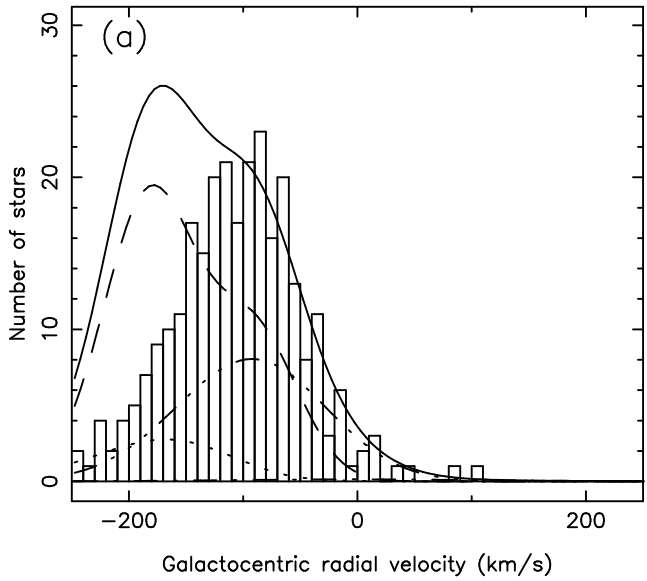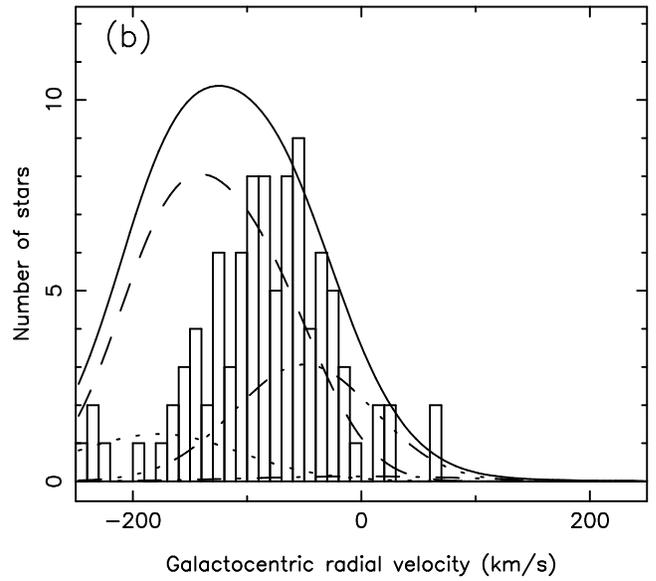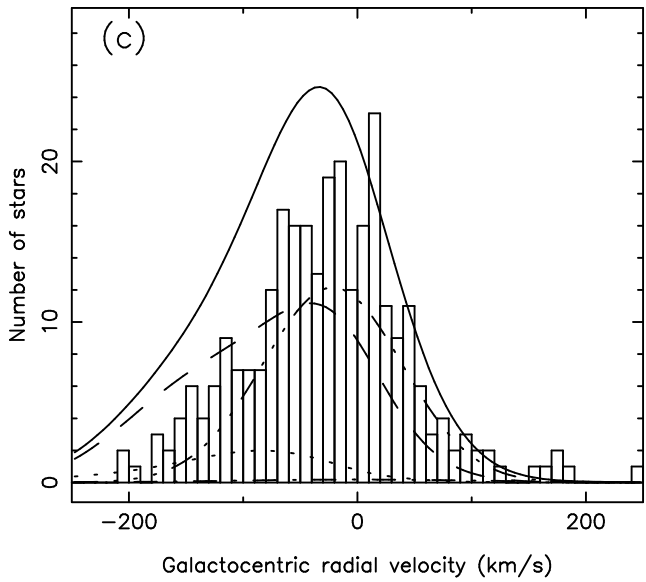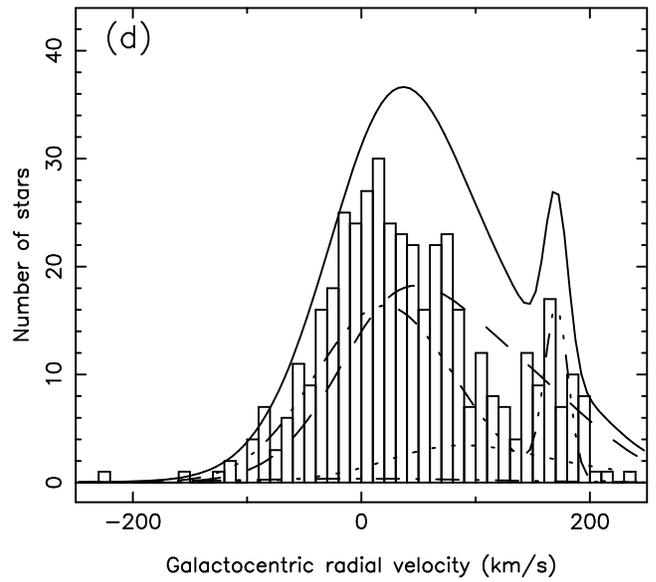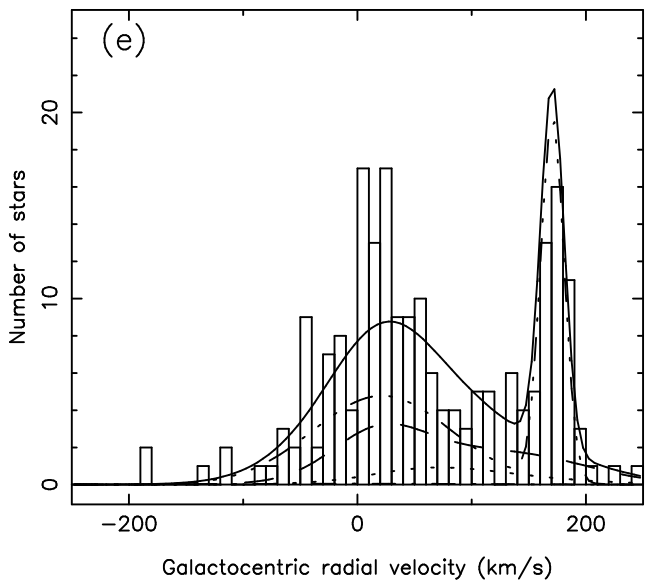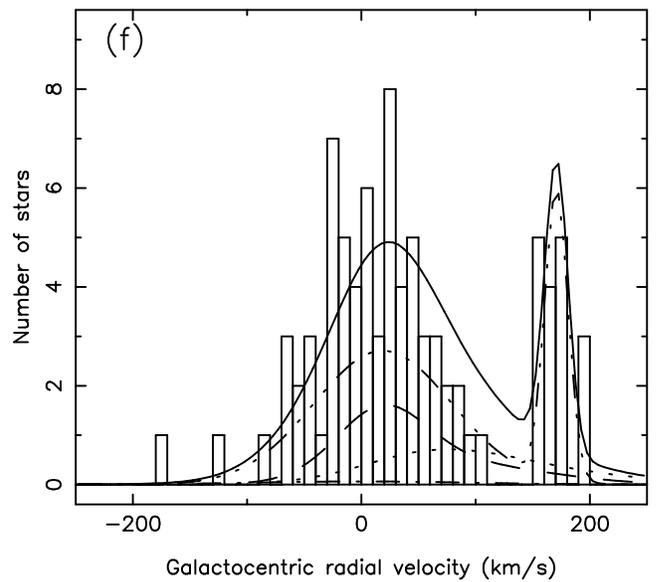

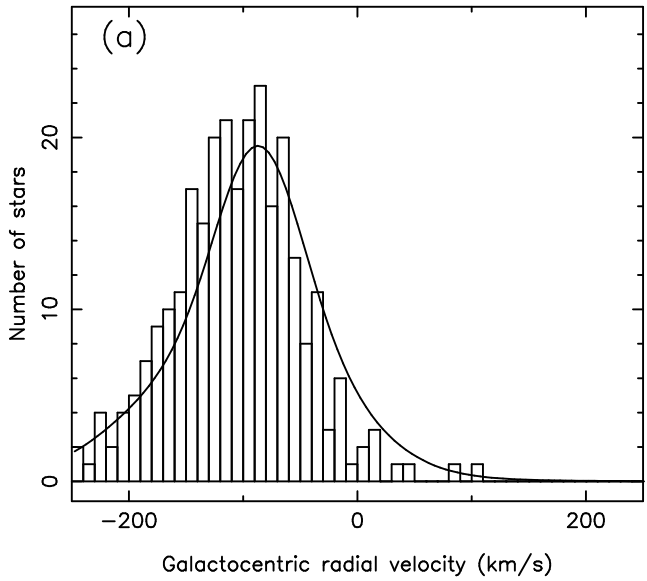
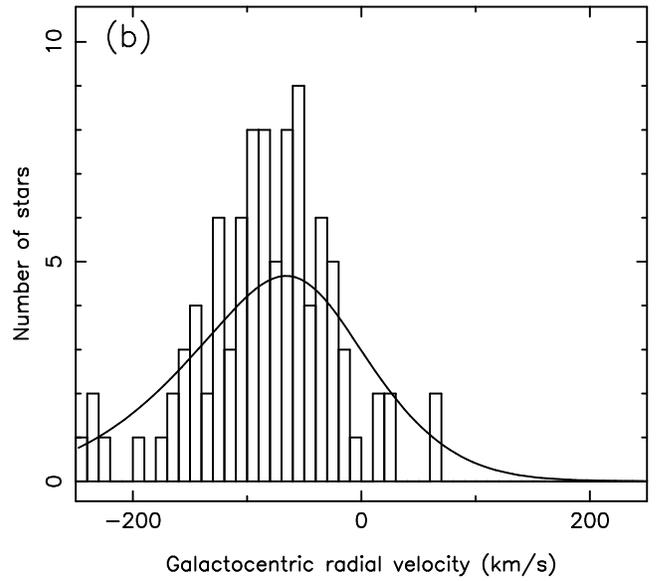
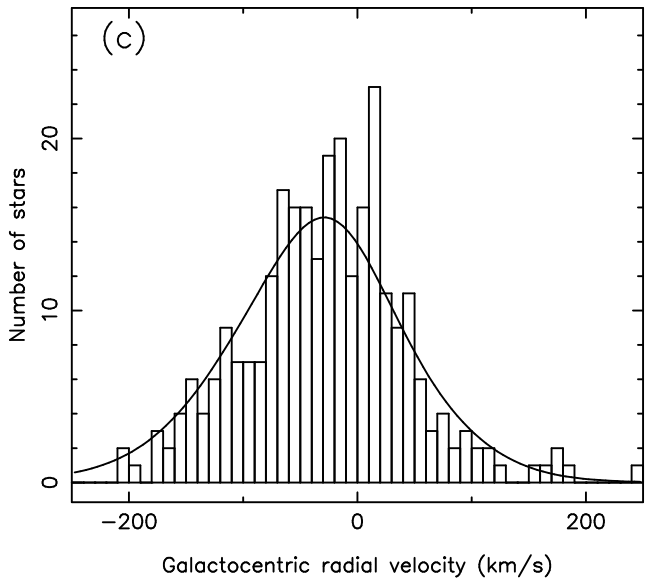
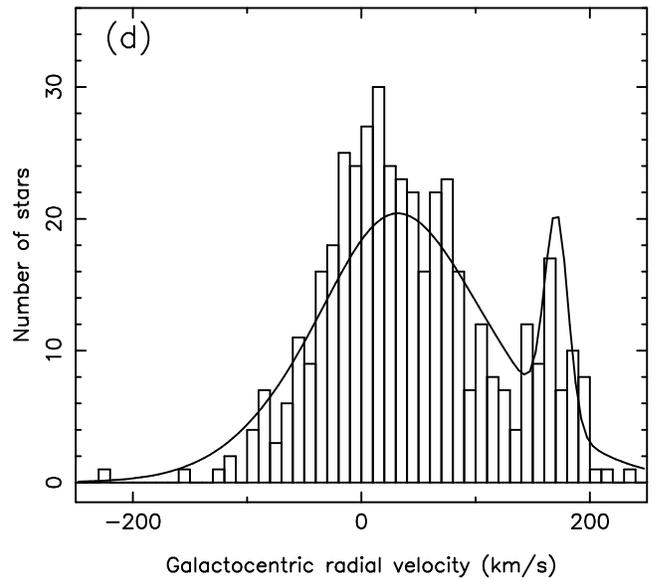
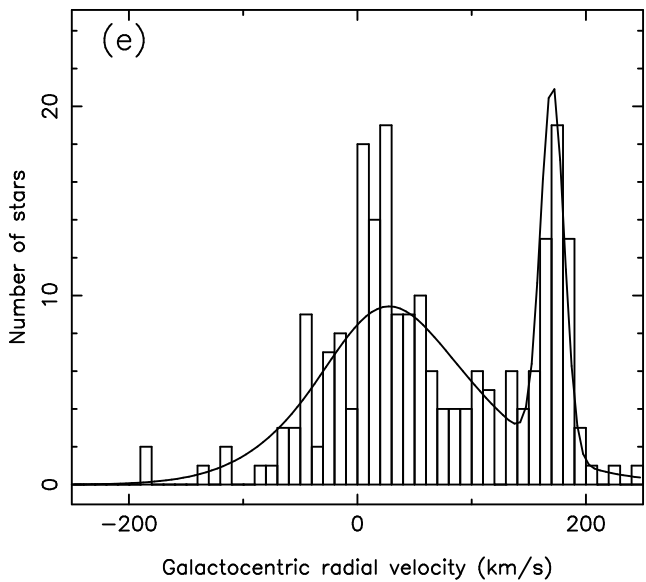
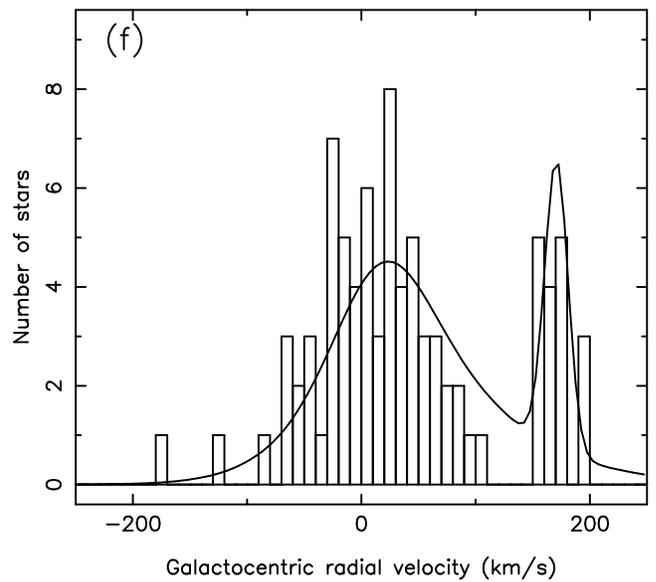

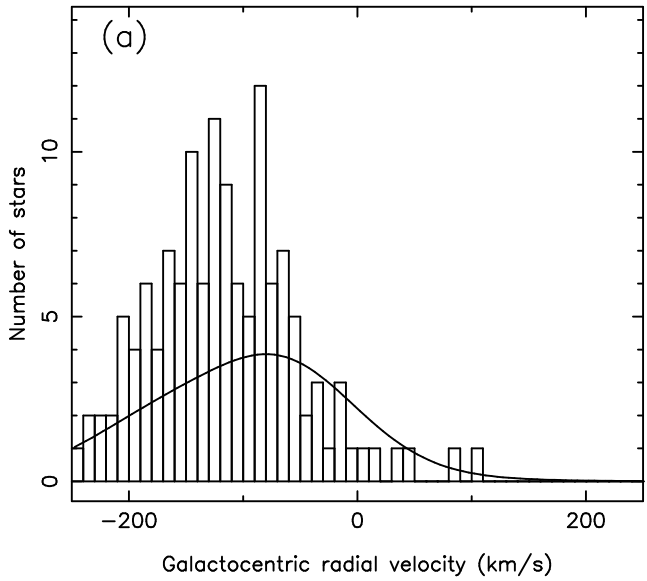
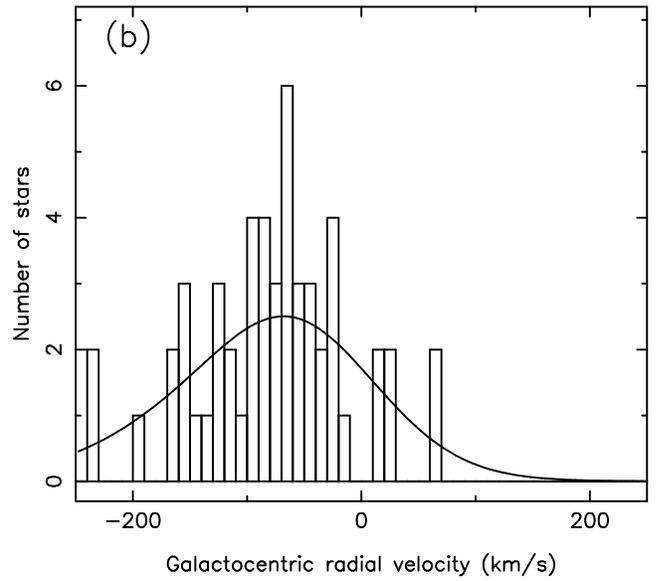
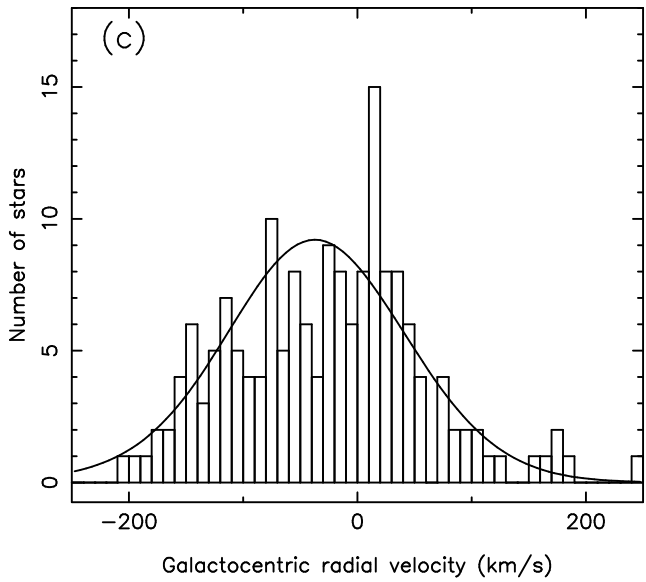
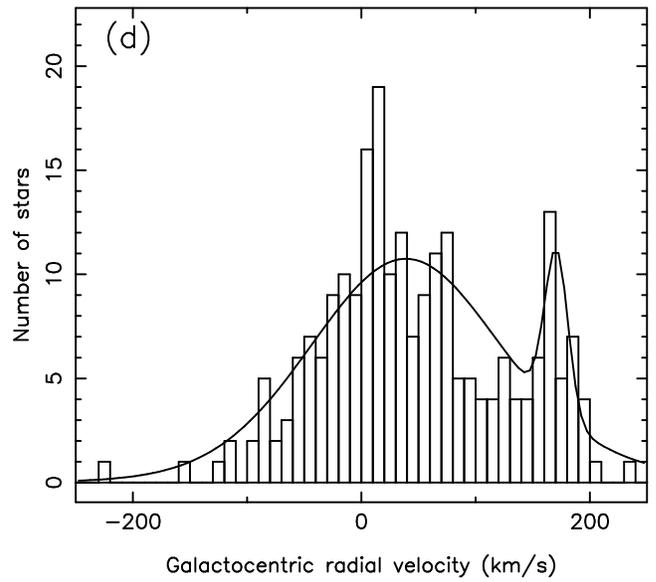
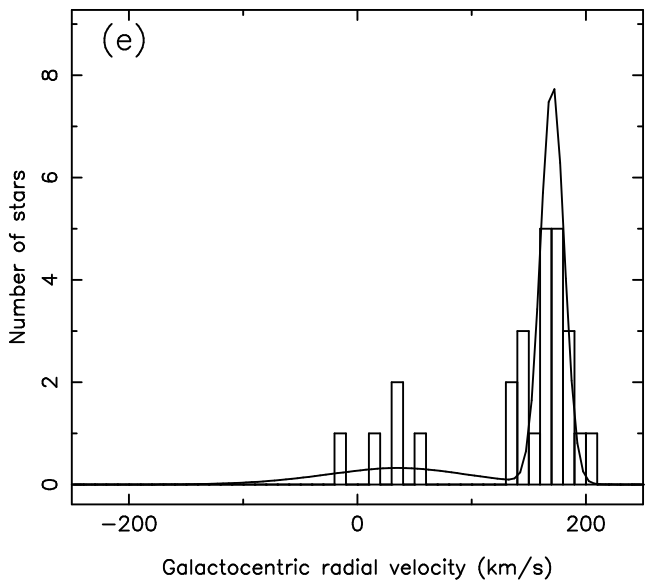
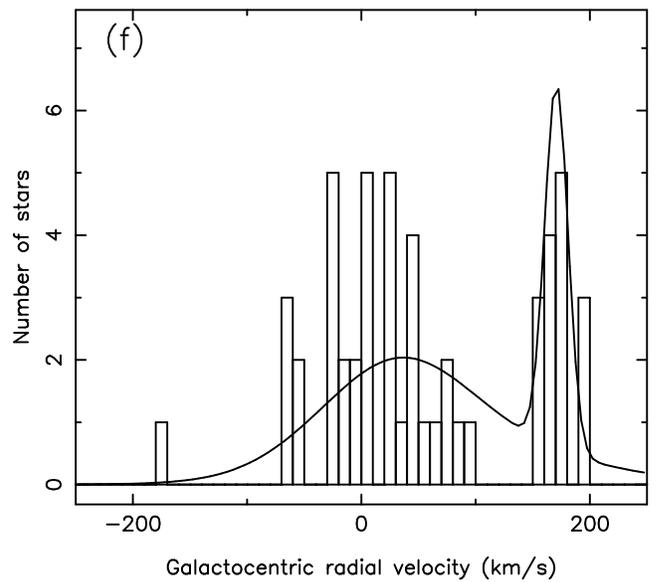

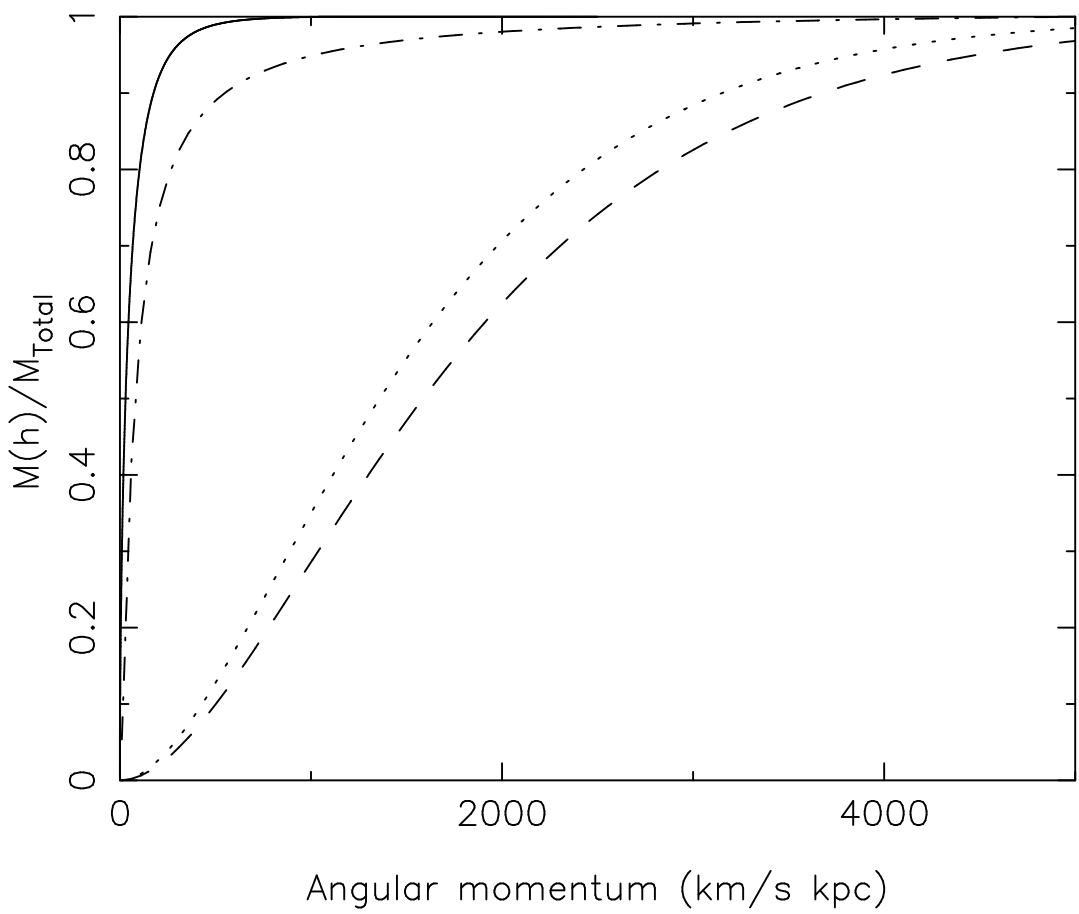

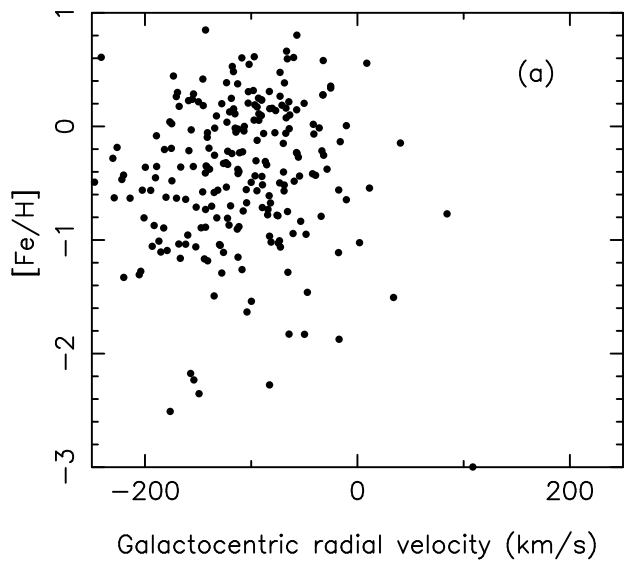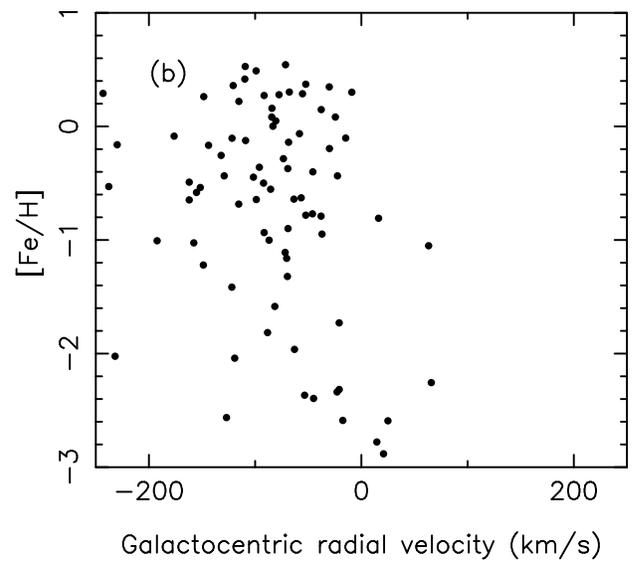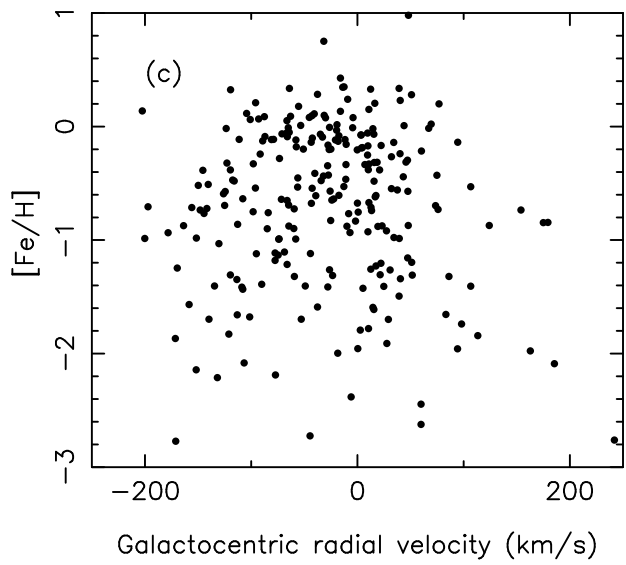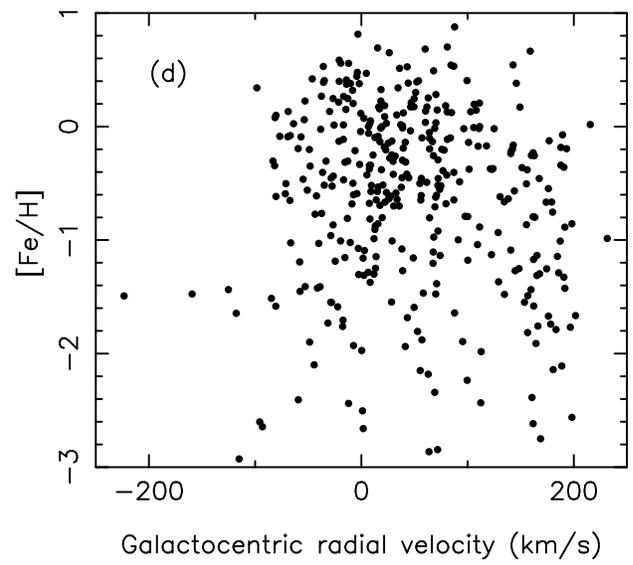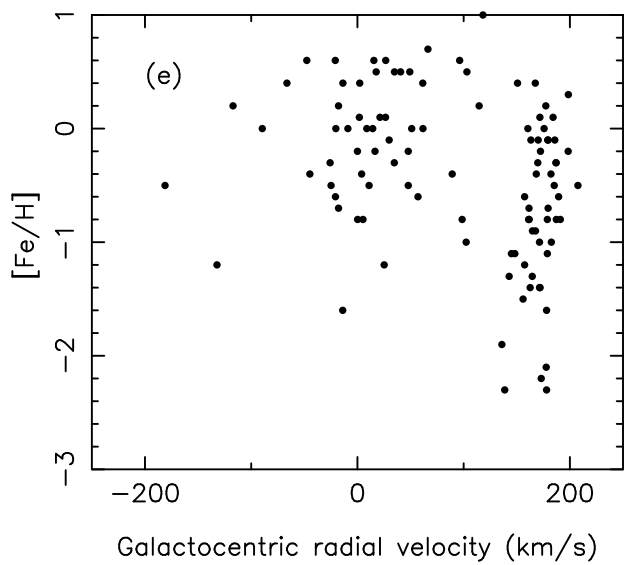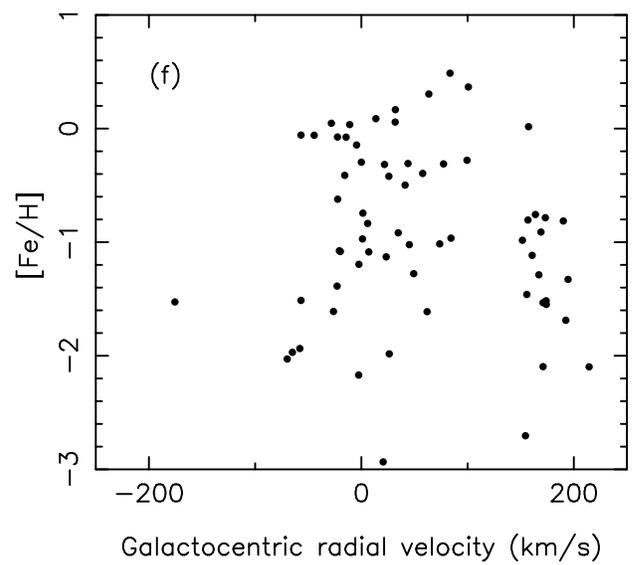

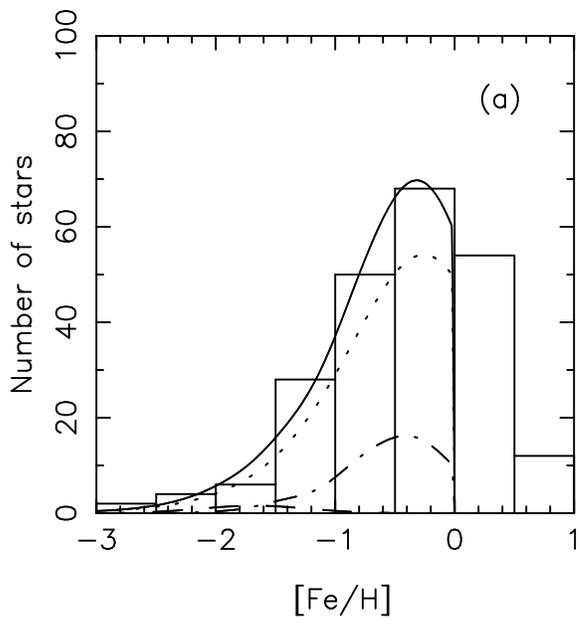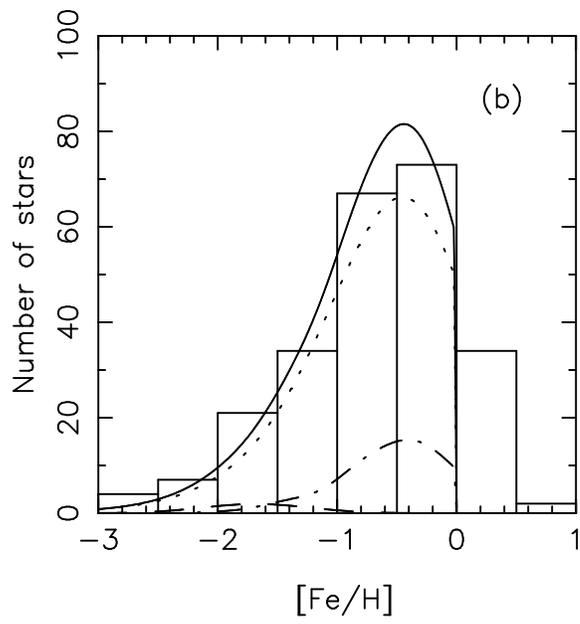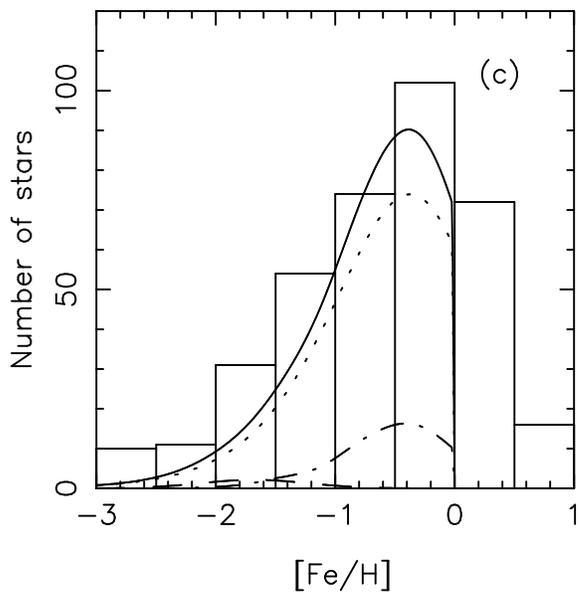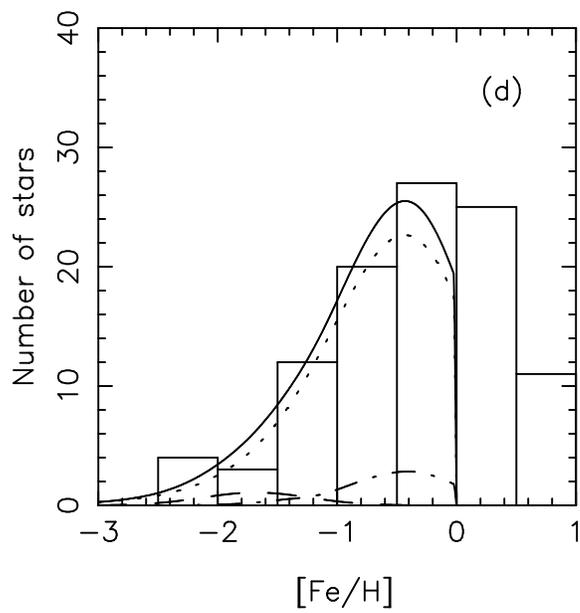

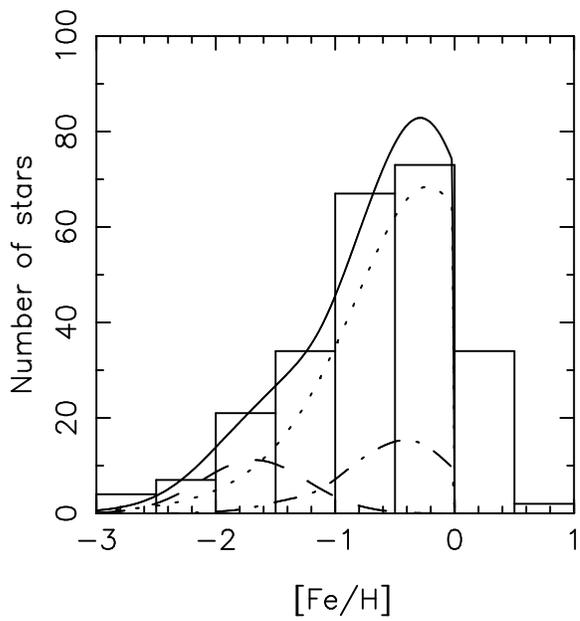

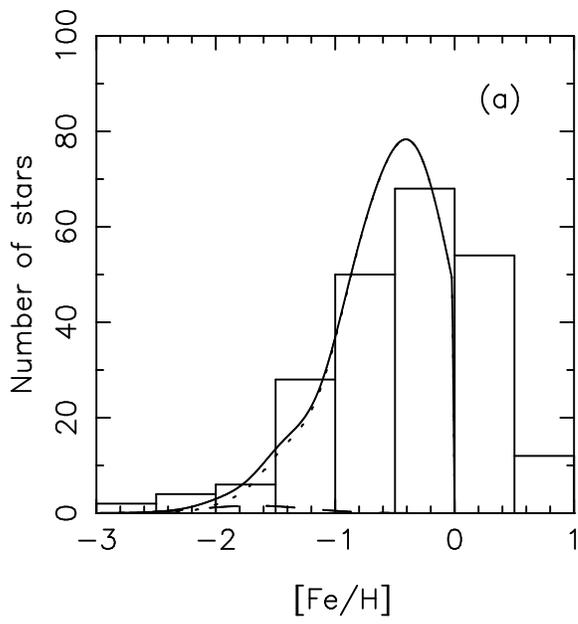
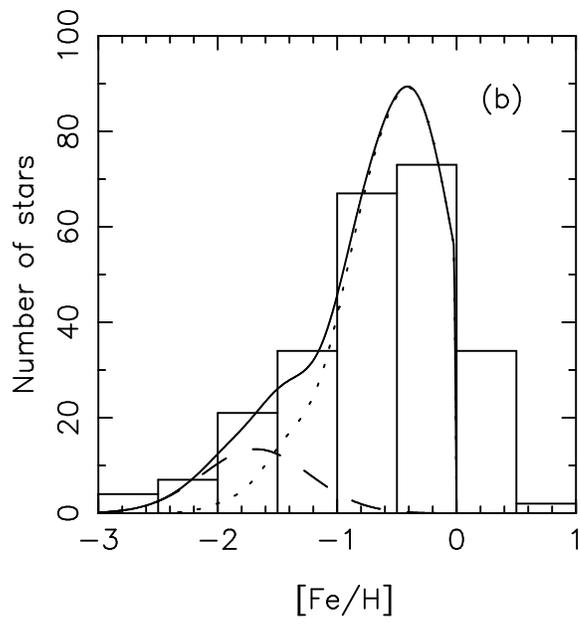
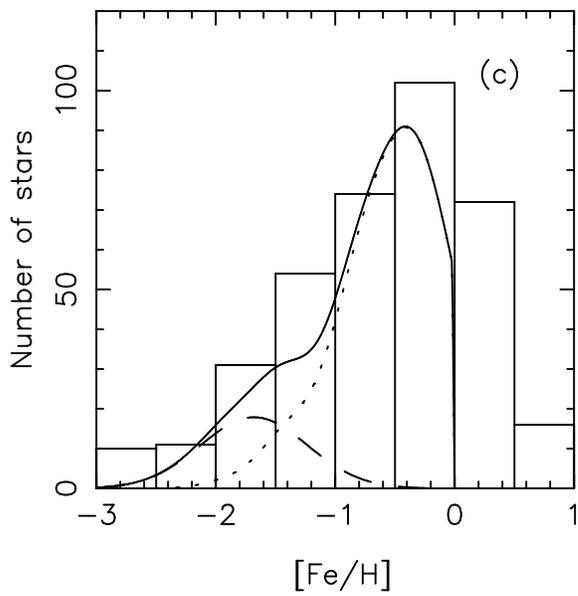
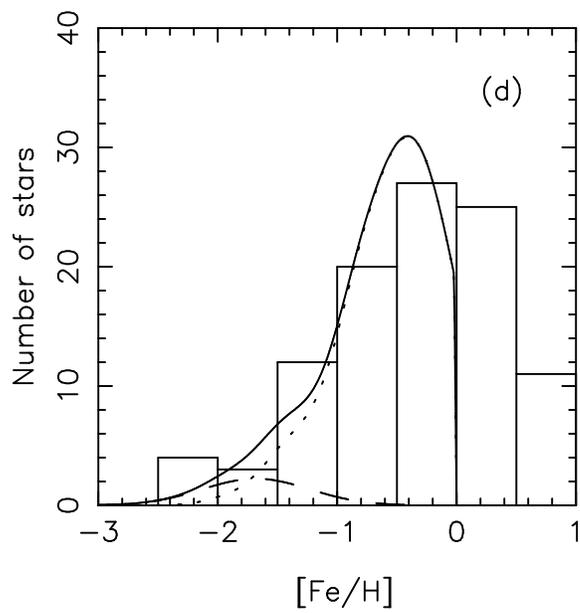